\documentclass[submission, Phys]{SciPost}

\usepackage{bm,amssymb,slashed,graphicx,multirow,soul,mathtools,xspace,array}  
\usepackage{float}   
\allowdisplaybreaks
\usepackage{ bbold }  
\usepackage{ifthen}
\usepackage{subcaption}
\usepackage{booktabs}
\graphicspath{{./figs/}}  

\providecommand{\doi}[1]{\href{http://dx.doi.org/#1}{doi:#1}}

\usepackage[labelfont=bf]{caption}

\newcommand{\beq}{\begin{equation} }
\newcommand{\eeq}{\end{equation}} 
\newcommand{\bi}{\begin{itemize} }
\newcommand{\ei}{\end{itemize} }

\definecolor{Red}{rgb}{1.,0.,0.}
\definecolor{Grn}{rgb}{0.,0.75,0.}
\definecolor{Blu}{rgb}{0.,0.,1.}
\definecolor{Purp}{rgb}{0.3,0.1,0.6}

\newcommand{\package}[2]{%
  \textsc{#1}\ifthenelse{\equal{#2}{}}{\xspace}{~#2\xspace}}
  \newcommand{\mlhad}[1][]{\package{MLhad}{#1}}
\newcommand{\pythia}[1][]{\package{Pythia}{#1}}
\newcommand{\herwig}[1][]{\package{Herwig}{#1}}
\newcommand{\sherpa}[1][]{\package{Sherpa}{#1}}
\newcommand{\madgraph}[1][]{\package{MadGraph}{#1}}

\newcommand{\pytorch}[1][]{\package{PyTorch}{#1}}

\newcommand{\scikitlearn}[1][]{\package{Scikit-learn}{#1}}

\newcommand{\lang}[2]{%
  \texttt{#1}\ifthenelse{\equal{#2}{}}{\xspace}{\texttt{#2}\xspace}}

\newcommand{\python}[1][]{\lang{Python}{#1}}

\newcommand{\bc}{{\bm c}}
\newcommand{\bx}{{\bm x}}
\newcommand{\by}{{\bm y}}
\newcommand{\bz}{{\bm z}}
\newcommand{\btheta}{{\bm \theta}}
\newcommand{\bphi}{{\bm \phi}}
\newcommand{\bpsi}{{\bm \psi}}

\hypersetup{
    colorlinks,
    linkcolor={red!50!black},
    citecolor={blue!50!black},
    urlcolor={blue!80!black}
}

\begin{document}

\begin{center}{\Large \textbf{
Modeling hadronization using machine learning\\
}}\end{center}

\begin{center}
Phil Ilten \textsuperscript{1$\dagger$},
Tony Menzo \textsuperscript{1$\star$},
Ahmed Youssef\textsuperscript{$1\ddagger$}, and
Jure Zupan\textsuperscript{$1\mathsection$}
\end{center}

\begin{center}
\textsuperscript{\bf 1} Department of Physics, University of Cincinnati, Cincinnati, Ohio 45221,USA
\\
${}^\dagger$ {\small \sf philten@cern.ch},
${}^\star$ {\small \sf menzoad@mail.uc.edu},
${}^\ddagger$ {\small \sf youssead@ucmail.uc.edu},
${}^\mathsection$ {\small \sf zupanje@ucmail.uc.edu},

\end{center}

\begin{center}
\today
\end{center}


\section*{Abstract}
{
We present the first steps in the development of a new class of hadronization models utilizing machine learning techniques. We successfully implement, validate, and train a conditional sliced-Wasserstein autoencoder to replicate the \pythia generated kinematic distributions of first-hadron emissions, when the Lund string model of hadronization  implemented in \pythia  is restricted to the emissions of pions only. The trained models are then used to generate the full hadronization chains, with an IR cutoff energy imposed externally. The hadron multiplicities and cumulative kinematic distributions are shown to match the \pythia generated ones. We also discuss possible future generalizations of our results.
}

\vspace{10pt}
\noindent\rule{\textwidth}{1pt}
\tableofcontents\thispagestyle{fancy}
\noindent\rule{\textwidth}{1pt}
\vspace{10pt}

\section{Introduction}
\label{sec:intro}

A typical particle physics Monte Carlo event generator factorizes into three distinct steps or blocks of code: (i) the generation of the hard process, (ii)  parton shower,  and (iii) hadronization (including color reconnections). The first two steps are perturbative in their nature, and thus under good theoretical control, with significant efforts devoted to improving the precision even further. The algorithmic challenges  are efficient sampling of final state particle configurations, and taming the factorial growth of the calculations with the increasing number of particles. The simulation of the hard matrix element is performed either by a specialized code, e.g.,  \madgraph~\cite{Alwall:2014hca}, which only calculates the hard process, or is directly included in complete event generators, such as \pythia~\cite{Sjostrand:2014zea}, \herwig~\cite{Bellm:2015jjp}, or \sherpa~\cite{Bothmann:2019yzt}, that also perform the parton showering.

In contradistinction, the hadronization step is inherently non-perturbative. One is therefore forced to resort to phenomenological models inspired by non-perturbative discriptions such as lattice QCD. The two main models used in simulating hadronization are the Lund string model~\cite{Andersson:1983ia,Andersson:1998tv,Ferreres-Sole:2018vgo} and clustering model~\cite{Field:1982dg,Gottschalk:1983fm,Webber:1983if}. In the string model, quark--anti-quark pairs are thought of being connected by a string, a flux tube of the strong force confined in the lateral direction. As the quark--anti-quark pair moves apart, the string breaks, creating new quark--anti-quark pairs in the process, resulting in the emission of hadrons. These emissions are performed iteratively, breaking the string either from the left or the right side, with the final step modified \textit{post hoc} in order to provide an emission similar to the previous steps. This model requires extra parameters to describe the hadrons' transverse momenta and heavy particle suppression, and has some challenges describing baryon production. Over ${\mathcal O}({20})$ parameters are required by the string model to describe the hadronization.

In the clustering model, gluons are forced to split into quark--anti-quark pairs at longer distances (lower energy). All quark--anti-quark pairs are grouped into color singlet combinations with a distance scale that depends only on the evolution step, and not the hard process step of the Monte Carlo even generation. Hadrons are emitted from these universally pre-confined clusters via a series of two-body decays until only physical hadrons remain. The model has fewer parameters and naturally generates hadron transverse momenta. However, the decays of massive clusters lead to phenomenological problems such as predicting heavy baryon distributions which do not match data well.

Machine Learning (ML) techniques offer the possibility to build alternatives to the above two models of hadronization. Such ML models could be directly built from data and provide insights into the current phenomenological models. While ML techniques have recently entered into the development of event generators, through adaptive integration~\cite{Bishara:2019iwh,Badger:2020uow,Gao:2020vdv,Gao:2020zvv,Chahrour:2021eiv,Winterhalder:2021ngy}, ML based fast detector or event simulations \cite{Matchev:2020tbw,Alanazi:2020klf,Nachman:2020fff,Stienen:2020gns,Butter:2020qhk,Backes:2020vka,Danziger:2021eeg,Butter:2021csz,Biro:2021zgm,Howard:2021pos,Quetant:2021hgi,Bieringer:2022cbs,Buhmann:2020pmy},  and model parameter tuning~\cite{Ilten:2016csi,Andreassen:2019nnm}, the application of ML to the problem of hadronization as the final step in the Monte Carlo pipeline is entirely new, to the best of our knowledge. The present manuscript represents a proof of principle that building a full fledged ML based hadronization framework is possible. 

In principle, both Generative Adversarial Networks (GANs)~\cite{radford2016unsupervised} and Variational Auto-Encoders (VAEs)~\cite{kingma2014autoencoding} have demonstrated the ability for ML to generate convincing physical observables such as photographs. Using these techniques for hadronization introduces three unique challenges: (i) producing sets of physical observables that vary in size (unlike a fixed number of pixels), ranging from ${\mathcal O}({1})$ to ${\mathcal O}({10^4})$; (ii) strictly conserving certain physical quantities, e.g., momentum and energy; and (iii) learning from limited training sets which only provide coarse-grain detail. In this paper we present an architecture based on conditional sliced-Wasserstein autoencoders (cSWAE) \cite{DBLP:journals/corr/abs-1804-01947,DBLP:journals/corr/abs-1711-01558}, that overcomes the above challenges. The resulting code, \mlhad, is publicly available, see Appendix~\ref{sec:App:A}. We demonstrate the capabilities of \mlhad by training it on specially prepared \pythia hadronization outputs with an explicit IR cut-off. 
To speed up the training we perform a transformation that captures the bulk of the energy dependence of the \pythia hadronization output. However, we also show that, if this transformation is not performed, the cSWAE can still reproduce the energy dependence and thus should be able to reproduce any additional energy dependence that may be present in the hadronization process realized in nature.  We expect that the first version of the cSWAE architecture presented here can be upgraded to eventually be trained directly on data.

The paper is structured as follows. In Section~\ref{sec:SWAEs} we introduce conditional sliced-Wasserstein autoencoders and describe how these can be used to reproduce the Lund string model of hadronization.  In Section~\ref{eq:sec:pythia:reproduce} we then compare the trained \mlhad models to the results of a simplified  \pythia hadronization model. Section~ \ref{sec:conclusions} contains our conclusions and a brief discussion of future directions. Appendix~\ref{sec:App:A} contains details about the publicly accessible \mlhad code, while Appendix~\ref{sec:app:sliced:Wasserstein} gives further details on the sliced-Wasserstein distance.

\section{Conditional SWAEs and hadronization}
\label{sec:SWAEs}

\subsection{The simplified Lund string hadronization model}
\label{sec:simplified:Lund}
\begin{figure}[t]
\centering
\includegraphics[width=0.5\textwidth, scale = 0.5]{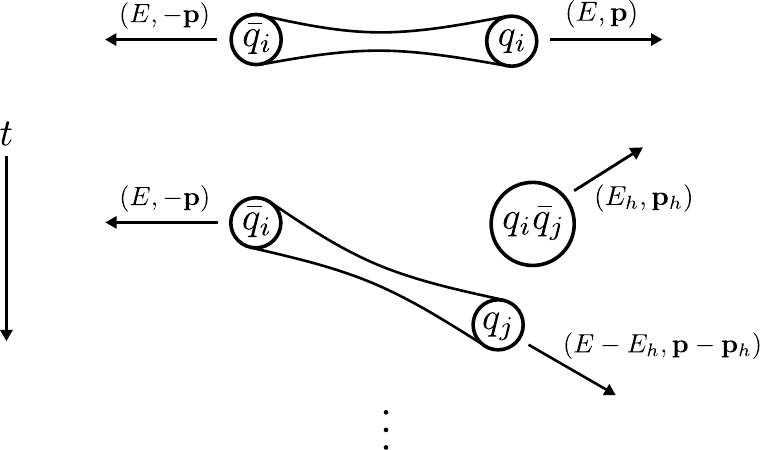}
\caption{\small The cartoon depiction of a single fragmentation event, viewed in the center-of-mass system of the initial string. The string connects the initial quark--anti-quark pair, $q_i\bar q_i$, each with energy $E$, moving back to back and carrying three-momenta $\pm \vec p$, respectively. In the hadronization event the string breaks and produces a hadron that is composed of valence quarks $q_i  \bar q_j$, and has energy $E_h$ and three momentum $\vec p_h$. Due to the flavor conservation the new string has as the new endpoints the $\bar q_i q_j$ quark--anti-quark pair, and the kinematics such that the energy and momentum are conserved. 
}
\label{hadron_cartoon}
\end{figure}

As the first step toward building a machine learning (ML) based simulator of  hadronization, we create a ML architecture that is able to reproduce a somewhat simplified Lund string model for hadronization. The physical process we want to describe is depicted in Fig.~\ref{hadron_cartoon}. It shows a $q_i \bar{q}_i$ fragmentation event in the center-of-mass frame in which the individual partons, each with flavor index $i$ and initial energy $E$, travel with equal and opposite momenta and are connected via a QCD string. String breaking produces a composite hadron $h \sim  q_i\bar{q}_j$ and a new $q_j\bar{q}_i$-string system depicted in the lower part of Fig.~\ref{hadron_cartoon}.\footnote{The depiction in  Fig.~\ref{hadron_cartoon} is for a string breaking occurring on the quark side. The string breaking on the anti-quark side produces similarly a hadron with quark composition $h\sim q_j \bar{q_i}$, and the new $q_i \bar{q}_j$-string.} The hadron $h$ is ejected with some energy and momentum $(E_h, \vec {p}_h)$, while the new string system has the energy and momentum $(2E-E_h , -\vec p_h)$, so that the total energy and momentum are conserved. 

After boosting to the center-of-mass frame of the new string, one has essentially the same initial state, a quark--anti-quark pair going back to back connected by a string, but with reduced energy $E'$ and a different quark flavor composition. Such fragmentation events stack one after the other and form a  fragmentation chain, one hadron emission at a time, until the entire energy of the initial two-parton system ($2E$) is  converted into hadrons. The end of the string used for each splitting is chosen at random. Until relatively low string energies of a few GeV, the selection of flavor and the kinematics of the hadron emission are taken to be independent processes. In the final stages of hadronization, when the string energy is close to the nonperturbative scale, the two processes, on the other hand, become intertwined. To simplify the problem, we therefore terminate fragmentation events at a center-of-mass string energy $E_{\rm cut}= 5$ GeV. We also consider a simplified string system which allows for $u$ and $d$ quarks as string ends, as well as their respective anti-quarks, and pions as final states. 

Note that each step in the above hadronization chain is independent from the previous one. A successful hadronization simulator therefore takes as the input the string energy $E$ (i.e., the energy of one of the endpoint quarks in the center-of-mass frame) as well as its flavor composition, and gives the flavor and kinematics of the hadron after first emission, $(E_h, \vec p_h)$. Repeating the first emission generates the full hadronization chain. Since $E_h^2={\vec p_h^2 +m_h^2}$, where $m_h$ is the hadron mass, the kinematics of the emission are fully described by specifying $\vec p_h$ and flavor of the created hadron $h$. We orient the coordinate system such that the $z$ axis is along the direction of the initial string, while the $x$ and $y$ coordinates are perpendicular to it. The transverse components of the $\vec p_h$ vector are given by
\begin{equation}
\label{eq:p:variables}
      p_x = p_T \cos \varphi, \hspace{0.2in} p_y = p_T \sin \varphi,
\end{equation}
where $p_T \equiv \sqrt{p_x^2 + p_y^2}$ and $\varphi$ is the polar angle. The string breaking and hadron emission are assumed to be axially symmetric in \pythia, i.e., independent of $\varphi$, and thus the problem of simulating the hadronization event reduces to a two variable problem of generating the $p_z$ and $p_T$ distributions for the first emission. 

\begin{figure}[t]
\centering
\includegraphics[width=0.9\textwidth, scale = 0.7]{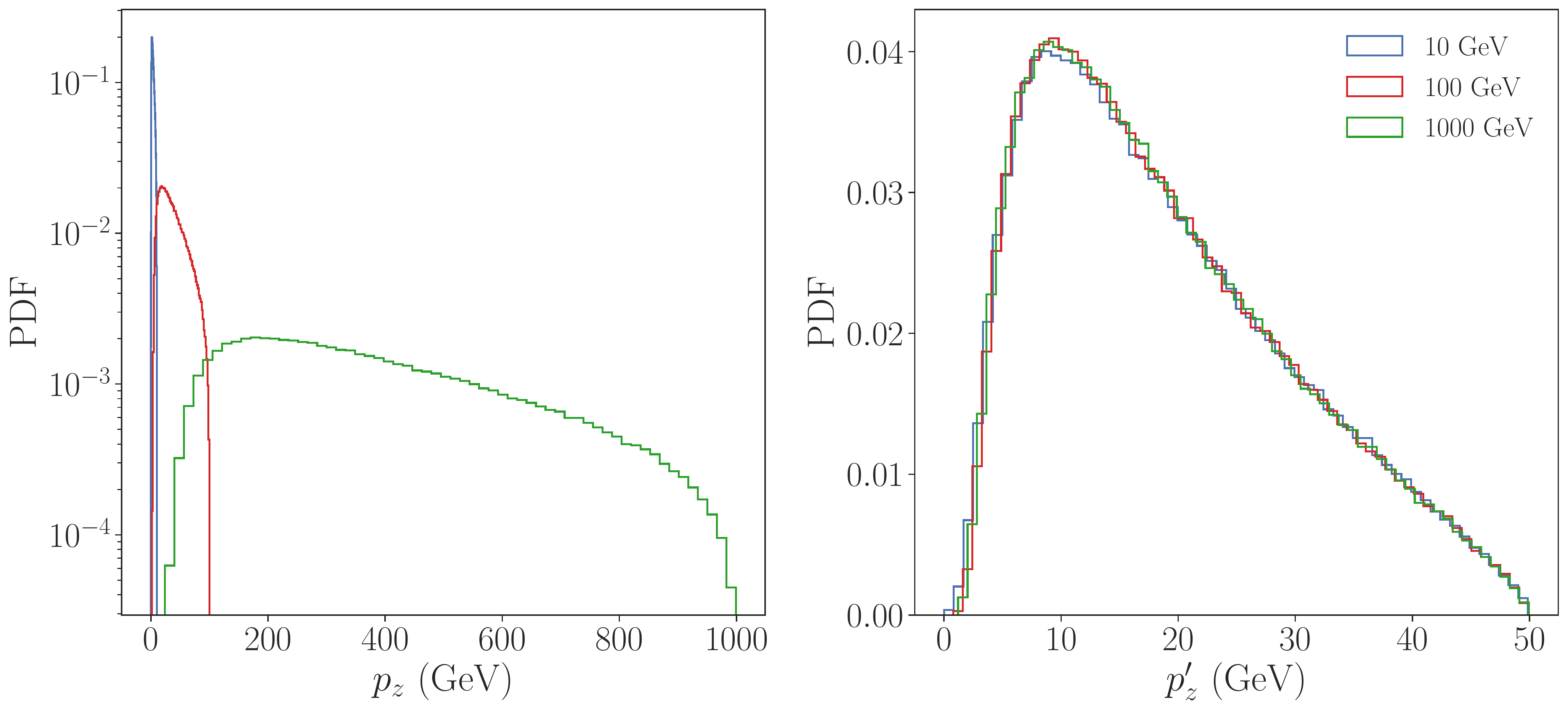}
\caption{The $p_z$ distributions (left) and the rescaled $p'_z$, Eq.~\eqref{eq:transform}, distributions (right) from \textsc{Pythia} hadronization events for the first-hadron emission with initial parton energies $E=10,100,1000$ GeV shown with blue, red, and green solid lines, respectively.}
\label{energy_independence}
\end{figure}

A special feature of the hadronization event and the chosen kinematic variables is the ability to render the $p_z$ kinematic distributions independent of the initial parton energy, $E$, through a simple rescaling transformation 
\begin{equation}
\label{eq:transform}
  p_z' \equiv E_{\rm ref} \frac{p}{E} ,
\end{equation}
where  $E$ is the energy of the quark in the center of mass for the initial string, and $E_{\rm ref}$ is a conveniently chosen reference energy that renders $p'$ dimensionful. In the rest of the paper we set $E_{\rm ref}=50$~GeV. The transformation of the $p_z$ distribution with respect to the initial parton energy $E$ can be seen in Fig.~\ref{energy_independence}.

The fragmentation process implemented in \textsc{Pythia} is constructed in momentum space as an iterative walk through production vertices. To do so a stochastic variable termed the longitudinal momentum fraction $z$ is defined, describing the fraction of longitudinal momentum taken away by the emitted hadron.\footnote{In Section~\ref{sec:cSWAE}, $\bz_i$ denote the latent-space variables. Despite similarity in notation there is no relation between the two variables.  } The probability distribution $f(z)$ from which $z$ is sampled is called the \textit{Lund left-right symmetric  scaling function (also  Lund sampling or fragmentation function)} and is given by
\begin{equation}\label{eq:lundFrag}
    f(z) \propto \frac{(1-z)^a}{z}\exp \left(-b \frac{m_{h,T}^2}{z} \right),
\end{equation}
where $m^2_{h,T} \equiv m^2_{h} + p_T^2 $ is the transverse mass, and the normalization prefactor is omitted for clarity. The phenomenological parameters $a, b$ are  chosen to match empirical data. The $p_T^2$ term in the transverse mass squared, $m^2_{h,T}$, captures the tunneling probability for a string breaking to occur away from the classical position of the string end, such that the  additional energy required for the transverse momentum kick can be released from the string. It leads to a correlation between transverse and longitudinal distributions of hadron momenta (in the  center-of-mass frame of the initial string),  i.e., the average value of $z$ increases with increasing $p_T$.  In the default implementation of the Lund model in \textsc{Pythia}, the  hadron $p_T$ distribution is assumed  to be Gaussian distributed, with average $\langle \vec p_T\rangle=0$, and a width $\sigma_0\sim{\mathcal O}(300\,\text{MeV})$, reflecting that its origin is an inherently quantum process occurring at the nonperturbative QCD scale.\footnote{The configurable \pythia parameter name is \texttt{StringPT:sigma}.}

The above basic setup of the Lund model becomes more involved when full complexity of the experimental data needs to be explained. Most of the ${\mathcal O}(20)$ parameters that give more flexibility to the  \textsc{Pythia} implementation of the Lund string model  are related to the differences in hadronizations of the light quarks compared to the strange, $c$ and $b$ quarks. For instance,   each quark flavor can in principle have a different $a$; in \textsc{Pythia} strange quarks are allowed to have different  values of $a$ than for $u$ and $d$ quarks, while for heavier $c$ and $b$ quarks the Lund fragmentation is also allowed to be multiplied by an extra $z$-dependent factor with new flavor-dependent parameters. Similarly, the $p_T$ distributions can deviate from the  Gaussian form. While this gives quite some flexibility to the hadronization model, it does have its own drawbacks. On one hand, the number of parameters to be tuned to data is already quite large. On the other hand, one may worry that the analytic form of the scaling function in Eq.~\eqref{eq:lundFrag}, while well motivated, is not flexible enough, with higher order corrections in $z$  potentially  becoming important, e.g., at low string energies. Generative ML models, such as the architecture that we introduce in the next section, can be used as effective tools to address both of these issues.

\subsection{The cSWAE architecture}
\label{sec:cSWAE}
The ML model of hadronization used here is based on the conditional sliced-Wasserstein Autoencoder (cSWAE) \cite{DBLP:journals/corr/abs-1804-01947,DBLP:journals/corr/abs-1711-01558}. The motivation for using cSWAE is two-fold, {\it i)} the flexibility of being able to use a wide variety of latent-space distributions and thus optimize the performance of the hadronization model, and {\it ii)} the ability to incorporate the energy dependence of hadronization through a two dimensional condition vector $\bc$. We expect the second feature to become most relevant once \mlhad is trained on experimental data, for which small breakings of the  energy independence exhibited by the Monte Carlo generated $p_z'$ data, Fig.~\ref{energy_independence}, may be anticipated. 

\begin{figure}[t]
\begin{center}
\includegraphics[width=0.6\textwidth]{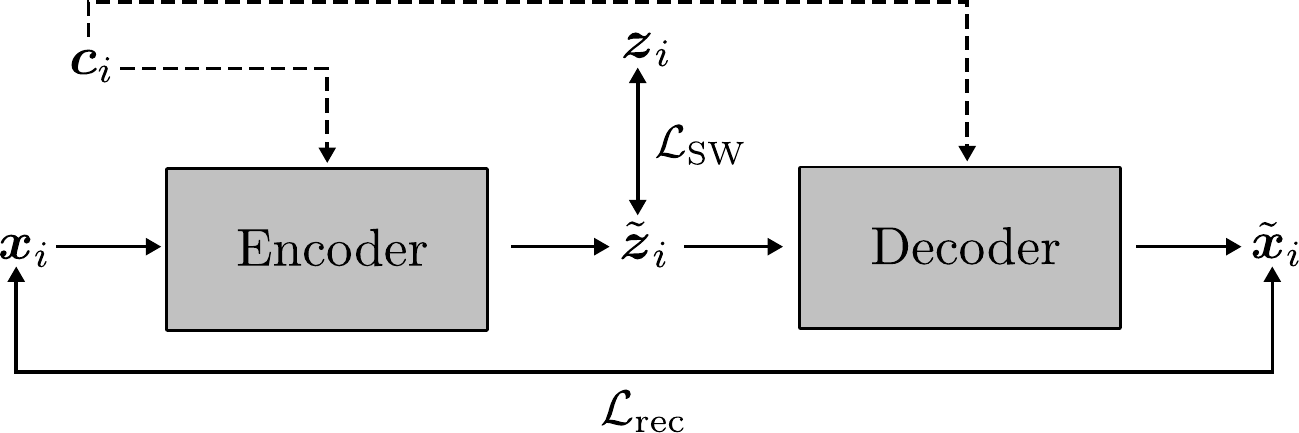}
\caption{The cSWAE architecture for simulating hadronization. The training data set are vectors  $\bx_i$ with sorted first emission hadron kinematics variables as their elements, either $ \bx_i=\{p_{z,k}'^{(i)}\}$ or ${\bm x}_i=\{p_{T,k}^{(i)}\}$. The ${\bm x}_i$ are inputs to the encoder, along with the pass-through condition vector ${\bm c}_i$, parametrizing the energy of the initial string. The decoder takes $ \tilde {\bm z}_i$ as inputs and generates the predicted hadron kinematics, either $\tilde \bx_i=\{\tilde p_{z,k}^{(i)}\}$ or $\tilde {\bx}_i=\{ \tilde p_{T,k}^{(i)}\}$. The sliced-Wasserstein-distance loss function, $\mathcal{L}_{SW}$, ensures that the latent-space vectors $\tilde {\bm z}_i$ follow the desired target distribution $\tilde {\bm z}_i\sim I(\tilde {\bm z}_i, {\bm c}_i)$. The reconstruction loss function, $\mathcal{L}_{\rm rec}$, minimizes the difference between input, ${\bm x}_i$, and output, $\tilde \bx_i$, first-emission hadron kinematics. The cSWAE architecture parameters are updated such that the sum $\mathcal{ L}_{\rm rec}+\mathcal{ L}_{\rm SW}$ is minimized.
}
\label{fig:cSWAE_architecure}
\end{center}
\end{figure}

The schematic of the cSWAE architecture is given in Fig.~\ref{fig:cSWAE_architecure}. It has two parts, the encoder and the decoder. The input data to the encoder are $N_e$ \pythia generated first-hadron emissions for a fixed initial string energy $E_i = 50$ GeV. In all of the numerical examples below we take $N_e=100$, so that the input is an $N_e$ dimensional vector ${\bm x}_i$
of either $p_{z,k}'^{(i)}$ or $p_{T,k}^{(i)}$, $k=1,\ldots, N_e$. That is, in this manuscript we apply cSWAE to the case where the $p_z'$ and $p_T$ distributions are uncorrelated and treat each of them separately. However, the architecture is flexible enough that  correlated 2D or higher dimensional distributions could also be used as inputs. 

The elements of the input vectors ${\bm x}_i$ are sorted, i.e., $p_{z,1}'^{(i)}\leq p_{z,2}'^{(i)}\leq \cdots \leq p_{z,N_e}'^{(i)}$ (and similarly for $p_{T,k}^{(i)}$).\footnote{ For 2D or higher dimensional problems the data would first be clustered in predefined 1D bins and then sorted within each bin.} The training dataset consists of $N_{\rm tr}$ such ${\bm x}_i$ input vectors, $i=1, \ldots, N_{\rm tr}$, and $N_{\rm val}$ $\by_j$ validation vectors, $j=1, \ldots, N_{\rm val}$, where typically $N_{\rm tr}$ is taken to be $N_{\rm tr}={\mathcal O}( 4000)$ and $N_{\rm val}$ an order of magnitude smaller. To summarize, the training and validation datasets are created by generating $N \equiv N_e (N_{\text{tr}}+N_{\text{val}}) = 4 \times 10^5$ \pythia first hadron emission events. The emission data ($p_z$ or $p_T$) is then partitioned randomly into $N_{\text{tr}}+N_{\rm val}$ vectors of length $N_e =100$. Finally, the elements in each vector are sorted from least to greatest.

The string energy $E_i$, or equivalently mass in the center-of-mass frame, is converted to a unit condition vector ${\bm c}_i=(\bar c_i, 1-\bar c_i)$ with $\bar c_i\in [0,1]$ a floating point number such that 
\beq \label{condition}
E_i= E_{\rm min}  \bar c_i + E_{\rm max} \big(1- \bar c_i\big), \qquad \text{and thus} \qquad \bar c_i = \frac{E_{\rm max} - E_i}{E_{\rm max} - E_{\rm min}},
\eeq
where $E_{\rm min}$ and $E_{\rm max}$ are the reference minimal and maximal energies. A good choice for  $E_{\rm max}$ is the maximal partonic collision energy in the simulation, while $E_{\rm min}$ can be taken to be the IR cutoff $E_{\rm cut}$. 

In general, the cSWAE allows for the initial string energy $E_i$ of each $\bx_i$ to be different (but the same for all the $N_e$ components of $\bx_i$). For the \textsc{Pythia} generated events the kinematic variable $p_z$ can be made $E$ independent through the transformation in Eq.~(\ref{eq:transform}) and thus $E_i$ can be set to a constant value, $E_i=50$ GeV. As a proof of principle we also show  in Section~\ref{sect:E_dep} that cSWAE models can be trained on $E$-dependent $\bx_i$. 

\begin{figure}[t]
\begin{center}
\includegraphics[width=0.6\textwidth, scale=0.5]{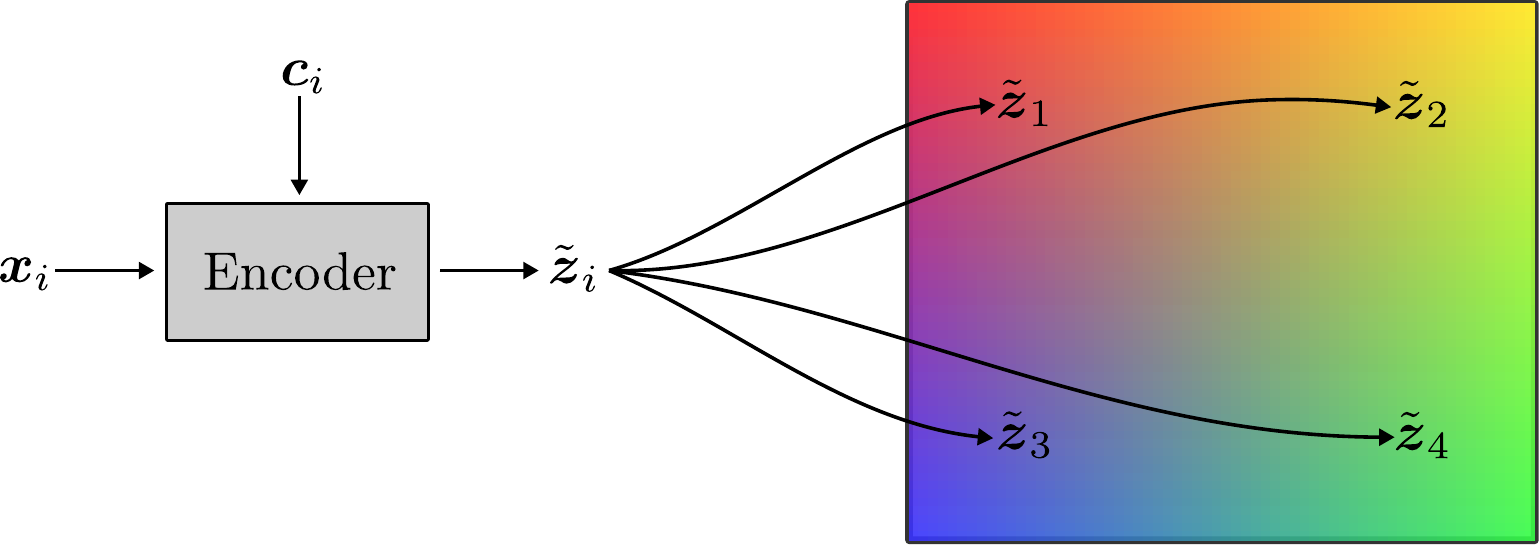}
\caption{Illustration of the role the conditional vector $\bc_i=\bc(E_i)$ plays in encoding the dependence of training data $\bx_i$ on the string energy $E_i$, by mapping the input data $\bx_i$ into different regions of the latent space, $\tilde \bz$.  After a sufficient amount of training, each area in the latent space corresponds to a different value of condition $\bc$. In \mlhad the condition vector $\bc$ is a continuous parameter and thus allows for interpolation to any given $\bc$ vector (string energy $E$). }
\label{fig:lat_mapping}
\end{center}
\end{figure}

The encoder $\phi$ takes as inputs the data vectors $\bx_i$ and labels $\bc_i$ and returns a latent-space vector $\tilde \bz_i=\bphi(\bx_i,\bc_i)$. Depending on the value of $\bc_i$ the encoder will transform $\bx_i$ to different regions in the latent space, as shown in the graphical representation of Fig.~\ref{fig:lat_mapping}.  
The dimension of the latent space, $d_z$, needed for the application to hadronization is anywhere from $d_z=2$ to $d_z=30$, see also Table~\ref{tab:SWAE_configs}. The latent-space vectors $\tilde \bz_i$ are trained to be distributed according to the target latent-space distribution, $\tilde \bz_i\sim I(\tilde \bz_i,\bc_i)$, which is ensured through the use of sliced-Wasserstein distance, $SW_p$, in the loss function.  In particular, the latent-space variable $\tilde \bz_i$ need not be normally distributed. We found that this feature translated to significant improvements in the performance of \mlhad.
With cSWAE one can choose a custom probability distribution  such that the encoding of the information about the first emission hadron kinematics leads to optimal results. This is the main practical difference between cSWAE and the conditional Variational Autoencoder (cVAE). The cVAE use KL-divergence in the loss function, which typically require that the latent-space variables follow simple distributions, such as a normal distribution. The cSWAE uses instead the sliced-Wasserstein distance, $SW_p$, see Appendix~\ref{sec:app:sliced:Wasserstein} for more details. This gives the architecture significantly more flexibility, as one can choose the latent-space distributions  to follow almost any  distribution, as long as it is sampleable (in particular, the analytic form of $I(\bz,\bc_i)$ is not required to exist).

\begin{figure}[t]
\begin{center}
\includegraphics[width=0.7\textwidth]{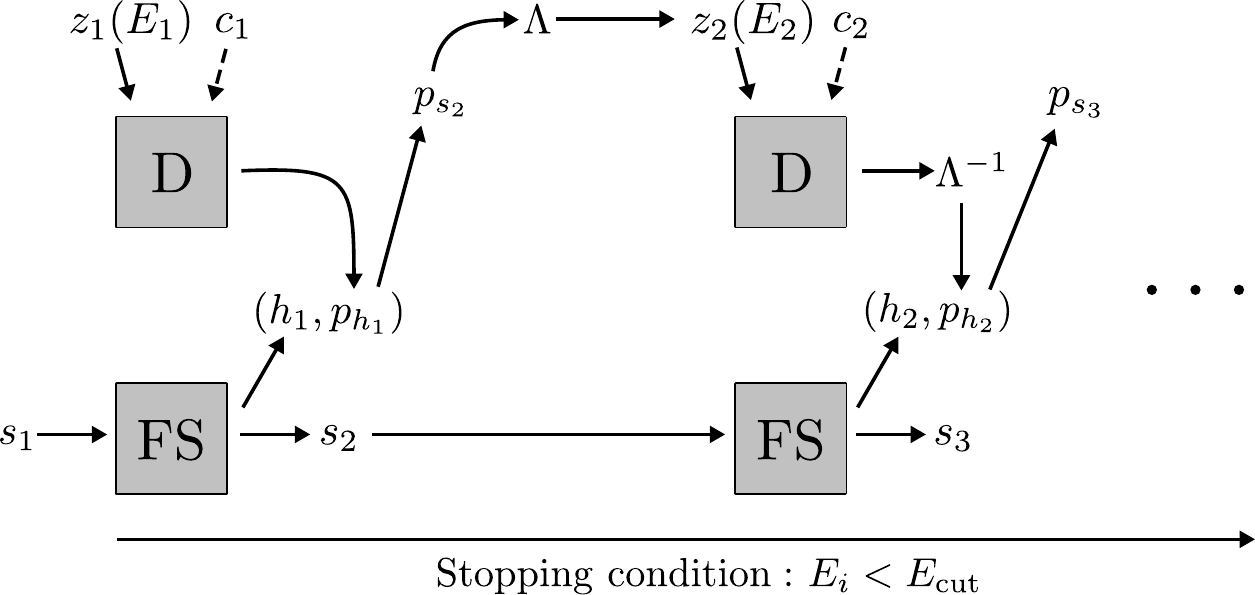}
\caption{An illustration of using \mlhad as a generator of hadronization chains. The decoder D is used as the generator of the hadron kinematics $(p'_{z,i}, p_{T,i})$, and thus also the four momentum of the new string fragment, $p_{s_{i+1}}$. The decoder takes as inputs the random variable $\bz_i$ from the latent space, and the conditional vector $\bc_{i}$ encoding the string fragment energy $E_i$ from the previous step. FS is the modified \textsc{Pythia} flavor selector, which takes as inputs the flavor composition of the string fragment from the previous step, $s_{i}$, and generates the flavor compositions of the new string fragment, $s_{i+1}$, and the flavor ID, $h_{i}$, of the emitted hadron. The generator chain is initialized in the first step with the energy of the initial string, giving the first conditional vector $\bc_1$ with its flavor ID $s_{1}$, and terminates when the energy of the string falls below a predetermined cut-off value, $E_{\rm cut}$. Before each hadron emission, the string fragments are boosted to its center-of-mass frame using a Lorentz transformation $\Lambda$.
}
\label{fig:frag_architecture}
\end{center}
\end{figure}

The decoder $\psi$ takes as inputs the condition vector $\bc_i$ and the latent-space vector $\tilde \bz_i$. It returns the reconstructed hadron kinematics $\tilde \bx_i=\bpsi(\bphi(\bx_i, \bc_i)) $,  where $\tilde \bx_i$ is the $N_e$ dimensional vector consisting of sorted kinematic variables, either $p_{z,k}'^{(i)}$ or $p_{T,k}^{(i)}$.  Through the minimization of the loss function \cite{DBLP:journals/corr/abs-1804-01947} 
\beq
\label{eq:Loss_fct}
\mathcal{L}(\psi, \phi) =\mathcal{L}_{\rm rec}+\mathcal{L}_{\rm SW},
\eeq
where
\begin{align}
\label{eq:Lrec}
\mathcal{L}_{\rm rec}=& \frac{1}{N_{\rm tr}} \sum_{i=1}^{N_{\rm tr}} \left[\frac{1}{Q} d_{2}^2(\bx_i, \bpsi (\bphi(\bx_i,\bc_i))) + d_{1}(\bx_i, \bpsi (\bphi(\bx_i,\bc_i))) \right], 
\\
\label{eq:LSW}
\mathcal{L}_{\rm SW}=& \frac{\lambda}{L N_{\rm tr}} \sum_{\ell=1}^L \sum_{i=1}^{N_{\rm tr}} d_{\rm SW}(\btheta_\ell \cdot {\bz}_{[i]_\ell}, \btheta_\ell \cdot \bphi(\bx_{[i]_\ell},\bc_i) ),
\end{align}
with $\bz_i\sim I(\bz_i,\bc_i)$, the training attempts to reproduce the training data  distribution $ \bx_i$ with the generated data distribution $\tilde \bx_i$, while the latent-space vectors $\tilde \bz_i$ follow the desired target distribution $\tilde \bz_i\sim I(\tilde \bz_i,\bc_i)$. 
The reconstruction loss ${\mathcal L}_{\rm rec}$ is a measure of the differences between the input, $\bx_i$, and generated kinematic vectors, $\tilde \bx_i$. It is the sum of two terms  for each of the 1D distributions that we consider,
\beq
\label{eq:dxi:yi2}
d_{2}^2(\bx_i, \bpsi (\bphi(\bx_i,\bc_i)))= 
\left\{
\begin{matrix}
 \sum_{k}\Big(p_{z,k}'^{(i)}-\tilde p_{z,k}'^{(i)}\Big)^2, & \text{for $p'_z$ distributions,}
 \\
 \sum_{k} \Big(p_{T,k}^{(i)}-\tilde p_{T,k}^{(i)}\Big)^2, & \text{for $p_T$ distributions,}
 \end{matrix}
 \right.
\eeq
\beq
\label{eq:dxi:yi}
d_{1}(\bx_i, \bpsi (\bphi(\bx_i,\bc_i)))= 
\left\{
\begin{matrix}
 \sum_{k}\big|p_{z,k}'^{(i)}-\tilde p_{z,k}'^{(i)}\big|, & \text{for $p'_z$ distributions,}
 \\
 \sum_{k} \big|p_{T,k}^{(i)}-\tilde p_{T,k}^{(i)}\big|, & \text{for $p_T$ distributions,}
 \end{matrix}
 \right.
\eeq
where $p_{z,k}'^{(i)}$ and $p_{T,k}^{(i)}$ are the components of the training-dataset vectors $\bx_i$, while $\tilde p_{z,k}'^{(i)}$ and $\tilde p_{T,k}^{(i)}$  are the components of the output vectors $\tilde \bx_i$. For the relative weight between the two terms in $\mathcal{L}_{\rm rec}$ we take $Q=1$\,GeV.

The second term in Eq.~\eqref{eq:Loss_fct}, $\mathcal{L}_{\rm SW}$,  is the implementation of the sliced-Wasserstein distance $SW_1$ between the distribution of latent-space vectors $\tilde \bz_i$ created by the encoder, and the target latent-space distribution $I(\bz_i,\bc_i)$. The vectors $\bz_i$ in Eq.~\eqref{eq:LSW} are randomly drawn from this target distribution, $\bz_i\sim I(\bz,\bc_i)$. The scalar products with the unit vectors $\btheta_l$, defining the $L$ slices, give the one dimensional projections of the latent-space distributions, for which the Wasserstein distances, $W_1$, are straightforward to compute. They are given simply by the average sum of the distances between the sorted data points, see Appendix~\ref{sec:app:sliced:Wasserstein} for further details. Note that for one dimensional latent space $SW_1=W_1$, and in the sum in Eq.~\eqref{eq:Loss_fct} one can set $L=1$.

The  algorithm for training the cSWAE  is as follows. Applying the encoder to the input data sample $\{\bx_1, .., \bx_{N_{\rm tr}}\}$ gives the latent-space vectors $\{\tilde \bz_1, .., \tilde \bz_{N_{\rm tr}}\}$. To compute the sliced-Wasserstein distance term, Eq.~\eqref{eq:LSW}, the unit vectors $\{\btheta_1,..,\btheta_L \}$ are randomly sampled from the $(d_z-1)$-dimensional unit sphere $\mathcal{S}^{d_z-1}$, while the $N_{\rm tr}$ latent-space vectors $\{\bz_1,\ldots,\bz_{N_{\rm tr}}\}$ are sampled from the target distribution, $\bz_i\sim I(\bz_i,\bc_i)$. For each $\btheta_\ell$, the scalar products $\btheta_\ell\cdot \tilde \bz_i=\btheta_l\cdot \bphi( \bx_i)$ and  $\btheta_\ell\cdot \bz_i$ are sorted in the following way. First the energy labels $c_i$ (and the corresponding $\tilde z_i$, $z_i$) are sorted into $N_c$ bins of increasing $c_i$ intervals with boundaries $\bar c_{[1]}<\bar c_{[2]}<\cdots< \bar c_{[N_c]}$. That is, the latent-space data are binned according to their energies, $E_i$, where the bins are chosen such that the distributions $I(\bz_i,\bc_i)$ do not have large dependence on $c_i$ within the bin. The generated and target $I(\bz_i,\bc_i)$ distributions are then compared within each energy bin. This is achieved by first sorting the scalar products of $\tilde \bz_i$ and $\bz_i$ with $\theta_\ell$ within each $c_i$ bin, and then combined into the lists $\{\btheta_\ell\cdot \tilde \bz_{[1]_\ell},\ldots, \btheta_\ell\cdot \tilde \bz_{[N_{\rm tr}]_\ell}\}$ and 
$\{ \btheta_\ell\cdot \bz_{[1]_\ell},\ldots,  \btheta_\ell\cdot \bz_{[N_{\rm tr}]_\ell}\}$, respectively. The SW loss function $\mathcal{L}_{\rm SW}$ in Eq.~\eqref{eq:LSW} is then the average over the latent space distances between the two sorted lists,
\beq
d_{\rm SW}(\btheta_\ell \cdot {\bz}_{[i]_\ell}, \btheta_\ell \cdot \bphi(\bx_{[i]_\ell})=\big| \btheta_\ell \cdot {\bz}_{[i]_\ell} - \btheta_\ell \cdot \bphi(\bx_{[i]_\ell})\big|,
\eeq
averaged also over all the $L$ slices and multiplied by the relative weight prefactor $\lambda$. The final step in the algorithm is applying the decoder to $\tilde \bz_i$, which gives $\{\tilde \bx_1,\ldots, \tilde \bx_{N_{\rm tr}}\}$. 
The distances between input dataset, $\{\bx_1, .., \bx_{N_{\rm tr}}\}$, and the generated sets $\{\tilde \bx_1,\ldots, \tilde \bx_{N_{\rm tr}}\}$ are then calculated using Eqs.~\eqref{eq:dxi:yi2} and ~\eqref{eq:dxi:yi}, giving the reconstruction loss function $\mathcal{L}_{\rm rec}$, Eq.~\eqref{eq:Lrec}. The decoder and encoder are updated in steps, trying to minimize the combined loss function, Eq.~\eqref{eq:Loss_fct}.
Overfitting is avoided by monitoring the value of loss function  when applied to the validation dataset, i.e., the loss function \eqref{eq:Loss_fct} with $\bx_i \to \by_i$, $N_{\rm tr}\to N_{\rm val}$.

\begin{figure}[t]
        \centering
        \includegraphics[width = 0.6\textwidth]{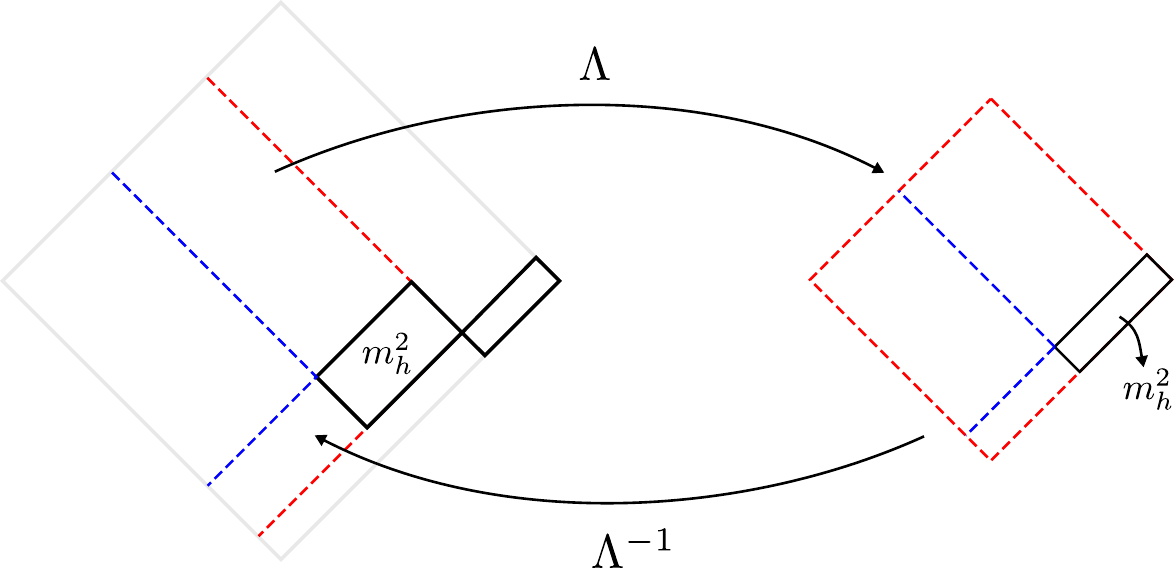}
        \caption{Illustrations of Lorentz boosting $(\Lambda)$ from the lab frame to the string center-of-mass frame. The red and blue lines denote the boundaries of the new string system's longitudinal momentum with the total area equal to the new string system's longitudinal momentum $E+p_z$. Each of the boxes can be considered as a `new' string system with scaled down energy. Perfectly square boxes indicate that we are in the center-of-mass frame. }
        \label{string_transform}
\end{figure}

Fig.~\ref{fig:frag_architecture} illustrates how the trained \mlhad decoder is used, along with the \pythia flavor selector, to generate the hadronization chain. Note, the full \pythia flavor selector is not needed here, but included to allow for subsequent development. The flavor selector takes as input the initial string flavor ID, $s_i$, and gives as the output the flavor ID of the emitted hadron, $h_{i}$, which also defines the flavor of the new string fragment, $s_{i+1}$. The \mlhad decoder takes as input the latent-space vector $\bz_i\sim I(\bz_i,\bc_i) $ sampled from the target distribution $I(\bz_i,\bc_i)$, where $\bc_i$ is the label encoding the center-of-mass energy of the string $s_i$, see Eq.~\eqref{condition}. The \mlhad decoder returns the $N_{e}$-dimensional vector with a list of possible momenta for the emitted hadron, $\tilde p_{z,k}'^{(i)}$ (or $\tilde p_{T, k}^{(i)}$). We randomly choose one of these as the actual hadron kinematics, and modify accordingly the kinematics of the remaining string fragment, $s_{i+1}$ , such that the energy and momentum are conserved.  The emitted hadron is boosted to the lab frame, and added to the list of emitted hadrons, while the new string is boosted to its rest frame, see Fig.~\ref{string_transform}. Its center-of-mass energy defines the label $\bc_{i+1}$ used as the input in the decoder for the next hadron emission.  These steps are repeated until the string energy in its rest frame reaches the IR cutoff energy $E_{\rm cut}$.

We have implemented the cSWAE architecture described above using \pytorch \cite{NEURIPS2019_9015}. The code can be accessed via a public repository, see Appendix~\ref{sec:App:A} for details.

\section{Reproducing the simplified \textsc{Pythia} fragmentation model}
\label{eq:sec:pythia:reproduce}
To demonstrate the viability and capability of the cSWAE based machine learning algorithm implemented in \mlhad, we reproduce the \pythia hadronization outputs. We analyze a $q_i \bar{q}_i$ hadronization event in the center-of-mass frame in which the individual partons, each with flavor index $i$ and initial energy $E$, travel with equal and opposite momenta producing a string between them. After the string breaks this produces a new string and the first emission hadron, see Section~\ref{sec:simplified:Lund} for more details.

While \mlhad treats all the hadron emissions on an equal footing, \pythia treats the first emission slightly differently; in the first emission $m_{T,h}$ in Eq.~\eqref{eq:lundFrag} is set to $m_h$ (i.e., $p_T=0$), while for all subsequent emissions $p_x$ and $p_y$ are sampled from a normal distribution with a width $\sigma_0$ (we set this tunable \pythia parameter to $\sigma_0 = 0.335 \text{\,GeV}$). Therefore, in training \mlhad we only aim to reproduce the $\pythia$ output {\em on average}, which is in line with the physical limitations of the problem, since one cannot trace in nature each individual emission in the hadronization event.

Our model is trained on kinematic distributions for transformed variables, $p'_z$, $p_T$,  Eq.~\eqref{eq:transform}, obtained from the \pythia  first emission events. With a uniformly sampled polar angle $\varphi$ in the transverse plane, these kinematic variables then completely define the phase space of the system through Eqs.~\eqref{eq:p:variables}, \eqref{eq:transform}. The \mlhad decoder is then used with a fixed shifted value transverse mass $m^2_{T,h}= m^2_h + \sigma^2$, with $\sigma =\sigma_0/\sqrt{2}$. This accounts for using only \pythia produced first emission data where $p_T = 0~\mathrm{GeV}$. For flavor selection we rely on \pythia's probabilistic model, and limit ourselves to light quarks, $u$, $d$ and only pions as the final state hadrons.

The independence of the distributions from the initial parton energy, see Fig.~\ref{energy_independence}, allows the cSWAE model to be trained on a dataset using an arbitrary initial parton energy, $E_{\rm ref}$, while the outputs of cSWAE hadronization generator can be transformed accordingly to obtain the distributions for any desired initial energy, $E$, using Eq.~\ref{eq:transform}. While in the \textsc{Pythia} output the complete energy dependence is already captured with the simple rescaling in Eq.~\eqref{eq:transform} we do not expect this to be entirely true for actual physical hadronization events realized in nature, for which subleading deviations from the scaling law in Eq.~\eqref{eq:transform} may be anticipated. In Section~\ref{sect:E_dep} we demonstrate that such corrections to the scaling law can be captured by the cSWAE architecture.

\subsection{First emission trained models}
\label{eq:trained:models}

\begin{table}[t]
\centering
\begin{tabular}{ c | c | c  c  c  c }
  \toprule
  Variable $\boldsymbol{x}$ & Target $\boldsymbol{z}$ & $t$ (epochs) & $d_z$ & $\lambda$ & $L$ \\
  \midrule
 \multirow{3}{*}{$p_z'$} & \pythia & 150 & 35 & 35 & 15 \\
  & Trapezoidal & 300 & 2 & 20 & 30 \\
  & Triangular & 150 & 2 & 30 & 25 \\
 \midrule
 \multirow{3}{*}{$p_T$} & \pythia & 100 & 20 & 30 & 30 \\
  & Skew-norm & 120 & 4 & 20 & 25 \\
  & Triangular & 120 & 4 & 15 & 25 \\
 \bottomrule
\end{tabular}
\caption{The cSWAE training configurations, see main text for details.}
\label{tab:SWAE_configs}
\end{table}

The cSWAE trained models differ according to the target latent-space distribution, $I(\bz, \bc)$, the dimension of the latent space $d_z$, training time $t$ (epochs), the value of the sliced-Wasserstein regularization parameter $\lambda$, and the number of slices $L$, as shown in Table~\ref{tab:SWAE_configs}. In all the cases we fix the string energy to be $E=50$ GeV. The first emissions for other string energies can be obtained by inverting the rescaling of the $p_z'$ distributions in Eq.~\eqref{eq:transform}, while $p_T$ distributions do not scale with $E$, although this is an assumption of the \pythia model. For \pythia generated $p_z'$ data we use the transverse pion mass $m^2_{T,\pi}= m^2_\pi + \sigma^2$, instead of the actual pion mass. Because of the different treatment of first and subsequent hadron emissions in \pythia, this choice for a pion mass will then reproduce the average \pythia hadronization results for full hadronization chains, as discussed in the beginning of Section~\ref{eq:sec:pythia:reproduce} and shown explicitly in Section~\ref{sec:hadronization:chain} below. 

A key feature of the SWAE algorithm and the sliced-Wasserstein loss is the ability to `push' the encoded latent space towards a target latent-space distribution. The choice of target distribution affects the total training time and the speed of kinematic data generation. Choosing a target latent-space distribution which is similar to the training data set distribution generally requires a fewer number of epochs to train the model to a specified accuracy compared to a target latent space which is dissimilar. This may come at a cost during the generation of kinematic data for hadronization events due to the generation of a large number of random variables obeying potentially complex probability distributions. 

\begin{figure}[t]
\centering
\includegraphics[width=0.9\textwidth, scale = 0.5]{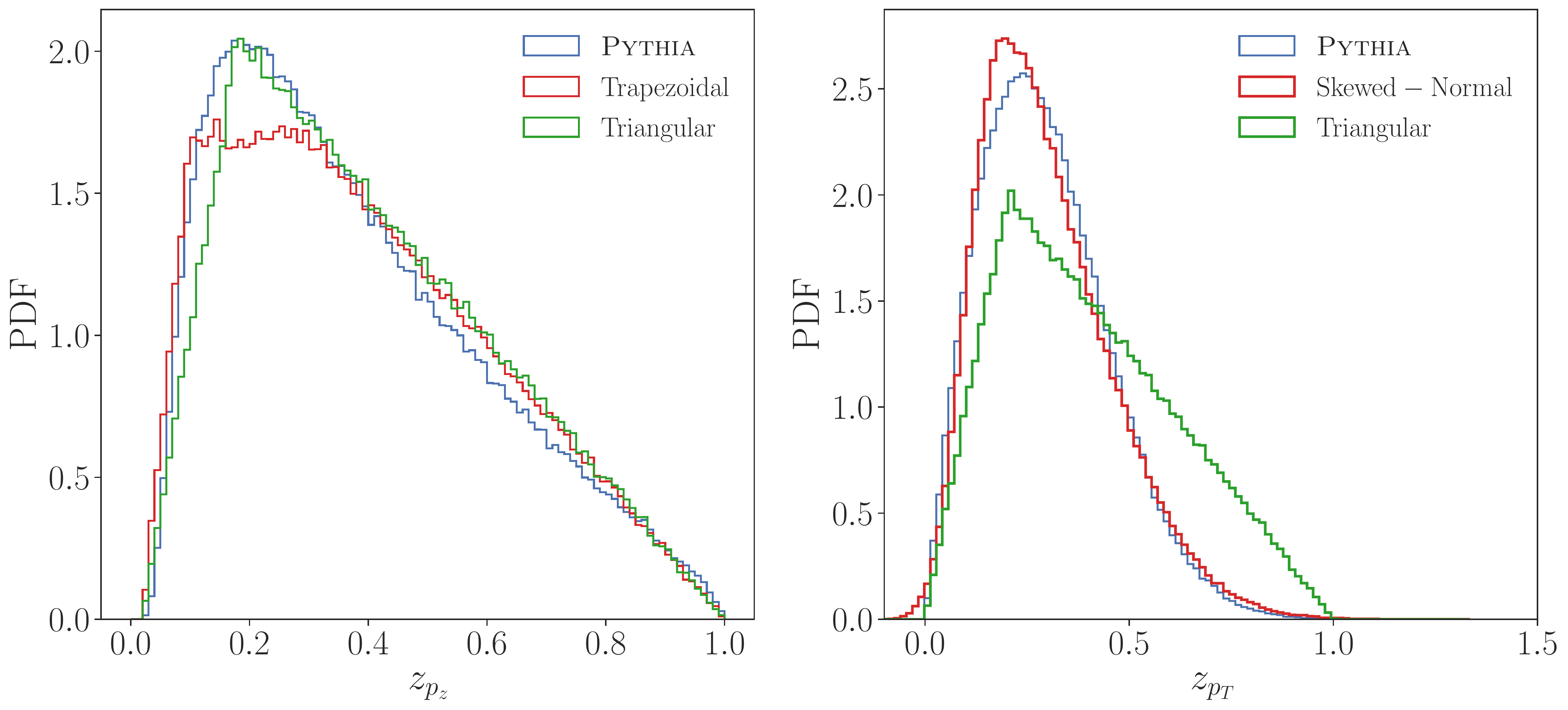}
\caption{Three choices for latent-space target distributions $I(\bz,\bc)$ for $p'_z$ inputs (left) and for $p_T$ inputs (right). See Appendix \ref{latent_dist} for more details.}
\label{pz_pT_latent}
\end{figure}

\begin{figure}[t]
\centering
\includegraphics[width=1.0\textwidth, scale = 0.5]{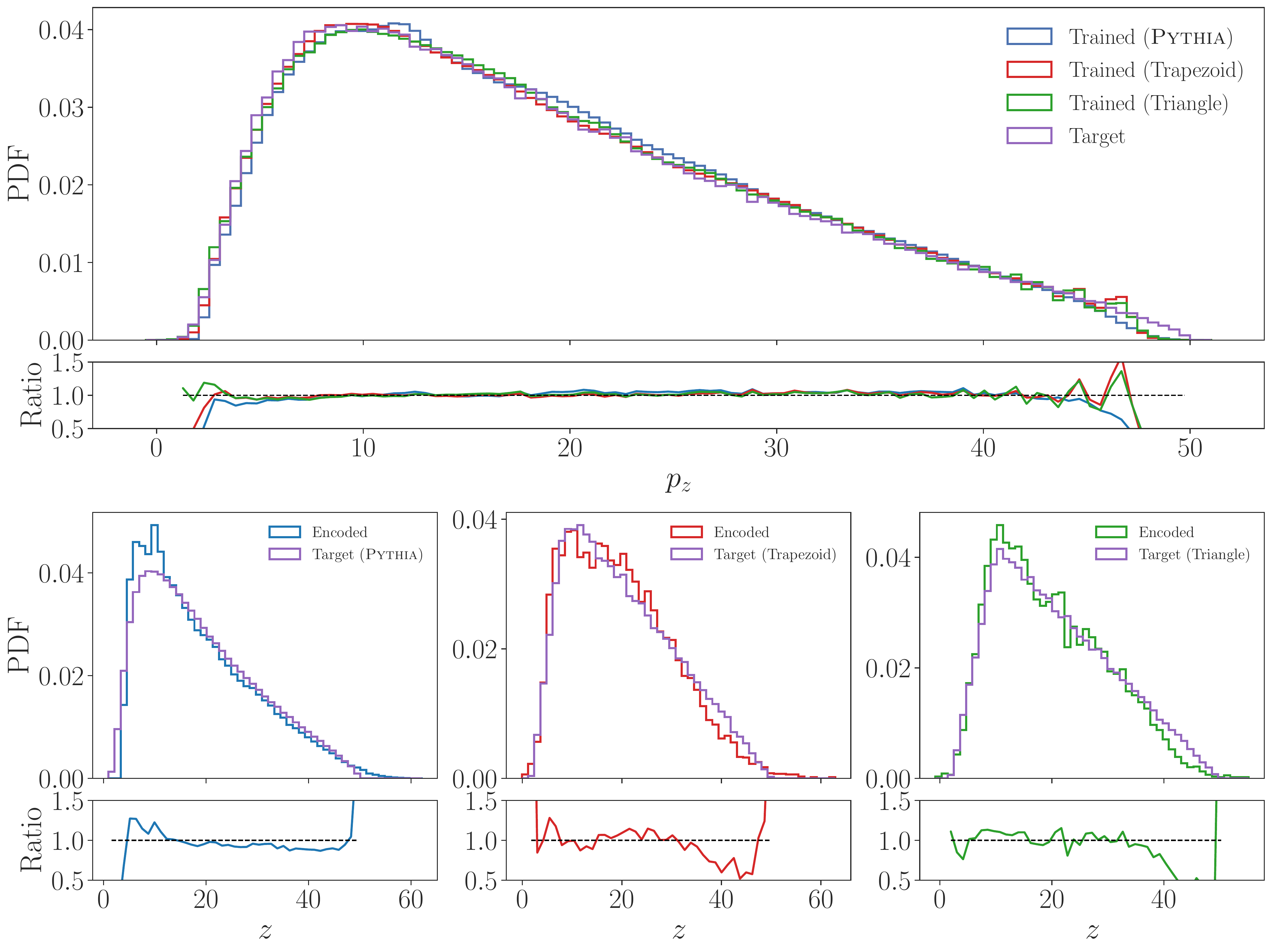}
\caption{Top: the \mlhad generated $p_z$ distributions for first-hadron emission from a string with an energy $E=50$ GeV, using three different latent-space distributions, \textsc{Pythia} (blue), trapezoidal (red), and triangular (green), compared to the \textsc{Pythia} generated target distribution (purple), as well as the ratios of \mlhad generated to \pythia generated distributions. Bottom: the comparison of the trained and target latent-space distributions for the three cases.}
\label{pz_gen_grid}
\end{figure}

\begin{figure}[t]
\centering
\includegraphics[width=1.0\textwidth, scale = 0.5]{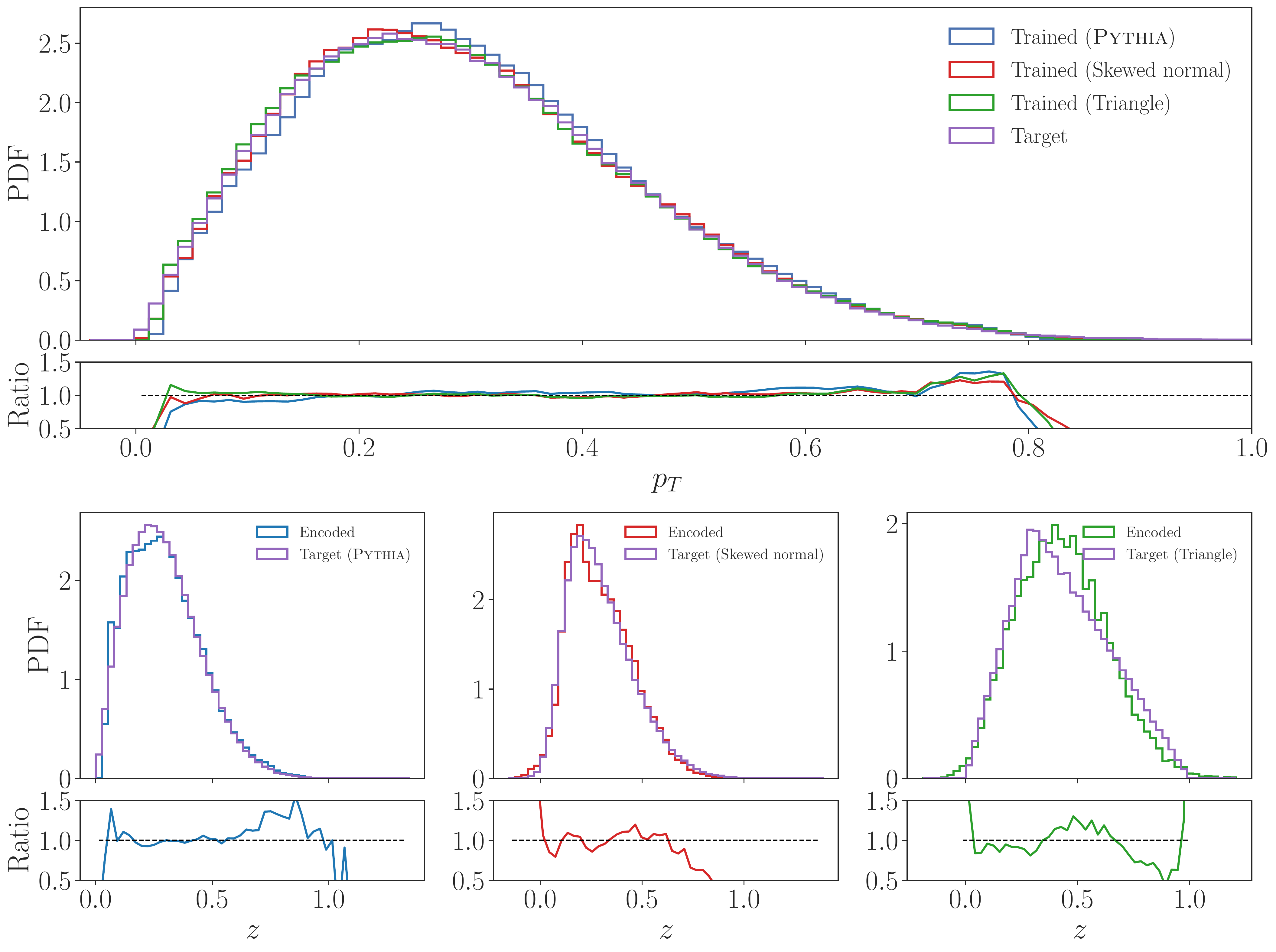}
\caption{Top: the \mlhad generated $p_T$ distributions for first-hadron emission using three different latent-space distributions, \textsc{Pythia} (blue), skewed-normal (red), and triangular (green), compared to the \textsc{Pythia} generated target distribution (purple), as well as the ratios of \mlhad generated to \textsc{Pythia} generated distributions. Bottom: the comparison of the trained and target latent-space distributions for the three cases.}
\label{pT_gen_grid}
\end{figure}

We demonstrate this flexibility by training with multiple target latent-space distributions, see Fig.~\ref{pz_pT_latent}. A total of six models are trained, three for each kinematic variable $p'_z$ and $p_T$, with the results shown in Figs. \ref{pz_gen_grid} and \ref{pT_gen_grid}. Of the three models in each kinematic variable, one model is trained using a target latent-space distribution equivalent to the training set distribution, i.e., the \pythia generated distribution of $p'_z$ or $p_T$. The other two trained models have target latent-space distributions that are distinctly different from the training set distributions. For $p'_z$ we choose trapezoidal and triangular target latent distributions and for $p_T$ we choose a skewed normal and triangular target latent-space distributions. The latent-space distributions are shown in Fig.~\ref{pz_pT_latent}, while their analytic forms can be found in Appendix~\ref{latent_dist}. Regardless of the choice of the latent-space distribution, the trained and the target (prior) data distributions are in good agreement.

 The dimension of the latent space is a tunable discrete hyperparameter, taking values $d_z \in [2,35]$, see the fourth column in Table~\ref{tab:SWAE_configs}. The regularization parameter $\lambda$ controls the magnitude of the sliced-Wasserstein loss and determines its relative weight in the total loss function, see Eq.~\eqref{eq:Loss_fct}. In practice, the regularization parameter determines how closely the encoded latent-space distribution will agree with the chosen target latent-space distribution, $I(\bz,\bc)$. In our trained models the regularization parameter in the loss function Eq.~\eqref{eq:Loss_fct} takes values $\lambda \in [15,35]$, as listed in the fifth column in Table \ref{tab:SWAE_configs}. Larger values are chosen in models where the target latent-space distribution is similar to the training distribution. Large values of $\lambda$ effectively reduce the size of the explored manifold which maps decoder weight-configurations to values of the loss function (if we think of the decoder as a partition function and the loss function as a functional, large values of $\lambda$ place the decoder near a saddle-point configuration). This improves the convergence to the minimum of $\mathcal{L}_{\text{rec}}$, resulting in shorter training times. This can also be explained by describing the correlation between the minimization of $\mathcal{L}_\text{SW}$ and $\mathcal{L}_\text{rec}$.
 
 The number of slices or projections used in the sliced-Wasserstein loss is also a tunable hyperparameter taking values $L \in [15,30]$, as listed in the last column in Table \ref{tab:SWAE_configs}. Each model uses the kinematic data generated from $N=4\times 10^{5}$ first emission events partitioned into $N/N_e=4000$ $N_e$-dimensional vectors, where $80\%$ of the data is used as the training and $20\%$ as the validation set.
We use an initial learning rate value of $10^{-3}$ and utilize \pytorch's dynamic learning-rate scheduler to reduce the learning rate according to plateaus of the loss function during training.

\subsection{Labels and $E$ dependent distributions}
\label{sect:E_dep}
The trained models for the first-hadron emission presented in the previous section were all obtained for a fixed initial string energy, $E$. To reproduce the $\pythia$ model for the first-hadron emissions (for string fragments with energies above $E_{\rm cut}$)
this is all that is required. The $p_z'$ distributions for any string energy can be obtained from the reference value of $E=50$ GeV that we used in the training by performing the rescaling, cf. Eq.~\eqref{eq:transform} and Fig.~\ref{energy_independence}. The $p_T$ distributions for first emissions, on the other hand, are independent of the initial string energy. 

However, the above scaling behaviors are not expected to be exact in nature. For one,  at lower string energies the approximations in deriving the string Lund model are likely to fail - the quarks are not massless, and there may be couplings between $p_T$ and $m_h$ that are not captured by the simple transverse mass tunneling ansatz, Eq.~\eqref{eq:lundFrag}. Furthermore, the origin of $p_T$ distributions for first emissions is purely non-perturbative in nature, and thus the $E$ independence of $p_T$ distribution assumed in \pythia is not rooted in first principles. 

\begin{figure}[t]
\centering
\includegraphics[width=0.8\textwidth, scale = 0.5]{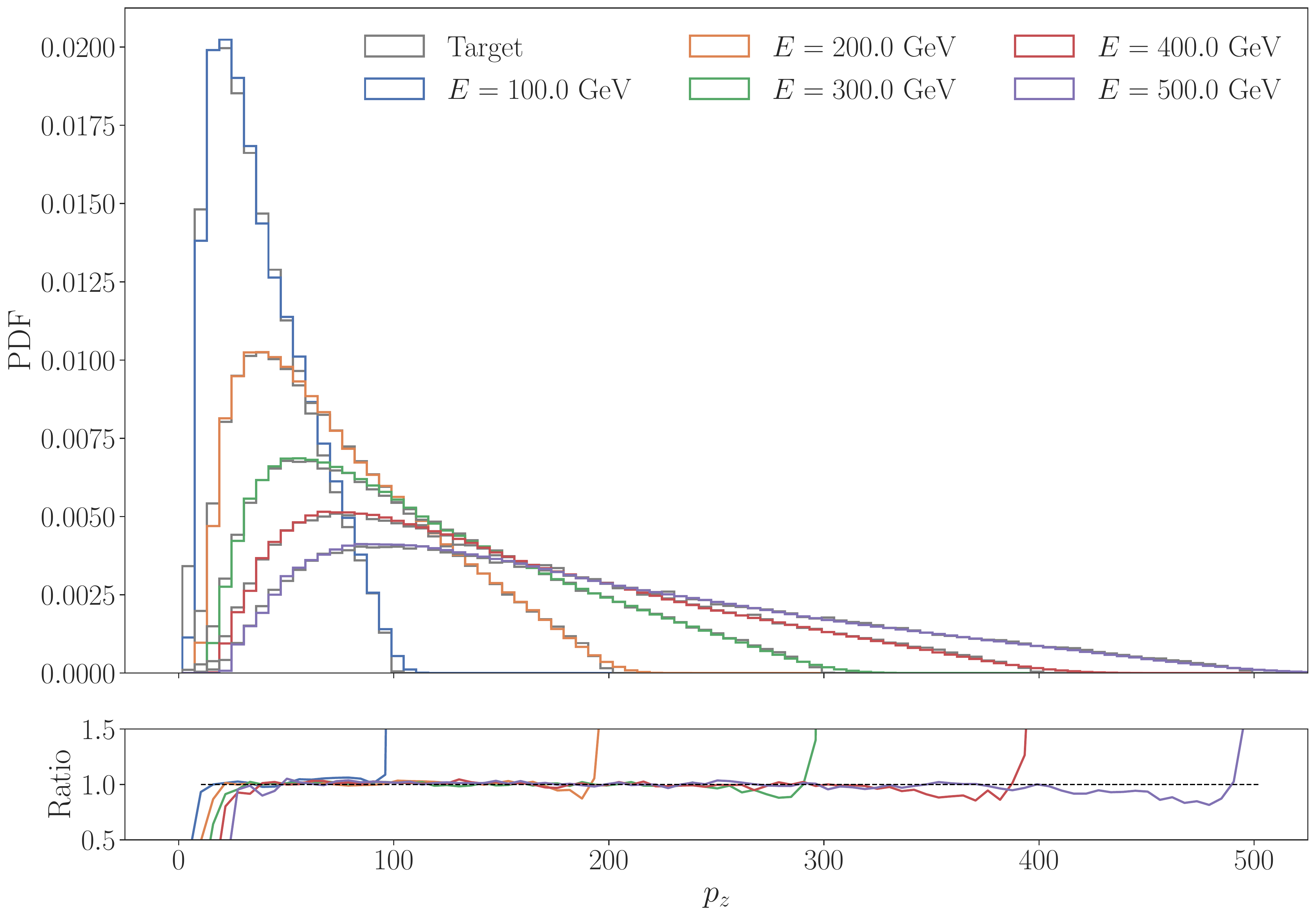}
\caption{The \mlhad generated  $p_z$ distributions for first-hadron emissions using the cSWAE model trained on data with string energies that differ from the ones used in the decoder, see text for details. The comparison with \pythia (black) demonstrates that \mlhad can faithfully interpolate to string energies never used in the training. }
\label{label_pz_gen}
\end{figure}

The \mlhad architecture is flexible enough to allow for the dependence of first emissions on the string energy, $E$. 
This is achieved by training the conditional SWAE on label-dependent datasets, which we demonstrate next. The training proceeds in a similar way as in the previous section, but now on a dataset comprising of first-hadron emissions for four distinct string energies, $E=\{5, 30, 700, 1000\}$\,GeV.\footnote{One could also have used emission data for continuous values of $E$, but binned finely enough in string energy values. We choose discrete string energies to demonstrate clearly that the cSWAE decoder can interpolate between the input labels.} Each $x_i$ input vector is therefore accompanied by one of the four discrete values for the two-dimensional vectors $\bc_i=(1-c_i, c_i)$ encoding the string energy through the label $c_i$ as defined in Eq.~(\ref{condition}), taking $E_{\rm min}=5$\,GeV and $E_{\rm max}=1000$\,GeV. 

The decoder in the trained cSWAE was then used to generate the first-hadron emissions at a different set of string energies, $E=\{100,200,300,400,500\}$\,GeV. Importantly, because the conditional vector is not discrete but rather depends on a continuous parameter defined between the minimum and maximum energies ($E_\text{min}, E_\text{max}$) the trained decoder is able to interpolate between labels (ones which the decoder has not trained on explicitly, see Fig.~\ref{fig:lat_mapping}) and rescale the kinematic distributions accordingly. 
This considerably increases the flexibility of generating training datasets as the user is able to choose the number of interpolation points which the model can use as anchors in generating data with a unique energy label.  The comparison of \mlhad and \pythia generated $p_z$ distributions for the first-hadron emissions is shown in Fig.~\ref{label_pz_gen}, demonstrating that \mlhad reproduces faithfully the \pythia results.

\subsection{Hadronization chain}
\label{sec:hadronization:chain}
As shown in the previous subsections the cSWAE trained models in \mlhad are able to accurately reproduce \pythia's first emission kinematics for a hadronized $q\bar{q}$ system in the center-of-mass frame of the string. In this section we show how well the \mlhad decoder  reproduces the full \pythia hadronization event. The implementation can be summarized as follows: from the initial string system, one string end is chosen randomly, while \pythia flavor selector is used to determine the flavor ID of the emitted hadron. Given the energy of the initial string end in the center-of-mass frame, $p'_z$ and $p_T$ are sampled using the corresponding cSWAE models. The $p'_z$ and $p_T$ of the emitted hadron are transformed to $p_x, p_y, p_z$ variables using Eqs. (\ref{eq:p:variables}) and (\ref{eq:transform}), and boosted to the lab frame. The string fragment is boosted to its center-of-mass frame, see Fig.~\ref{string_transform},  after which one repeats the hadron emission process 
until the string energy in the center of mass of the remaining string fragment falls below the IR cutoff, $E_{\text{cut}}$. The implemented fragmentation chain architecture is illustrated in Fig.~\ref{fig:frag_architecture}.  

\begin{figure}[t]
        \centering
        \includegraphics[width=0.8\textwidth]{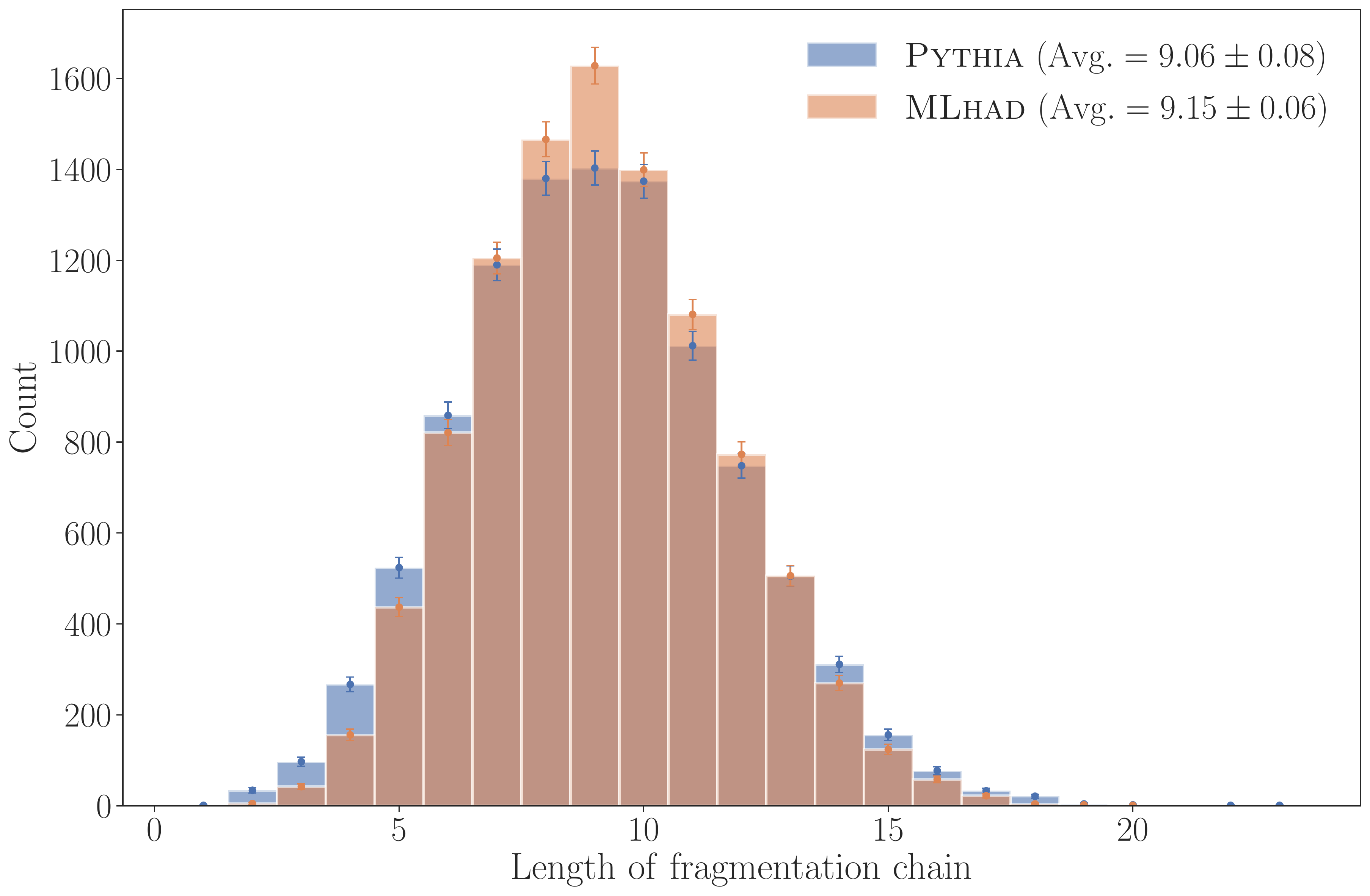}
        \caption{Comparison of the number of hadrons produced in the fragmentation chain of a single string for a sample of $10^4$ strings, compared between \pythia (blue) and \mlhad (red) generated hadronization events.}
        \label{fig:figure4}
\end{figure}

Fig.~\ref{fig:figure4} shows a comparison between the hadronization chain multiplicities obtained by \pythia (blue) and by the \mlhad model trained on first emission data (red). In both cases, starting from the initial string energy of $E=50$ GeV,  on average $9.1$ hadron emissions occur before the string fragment energy drops below the cutoff energy, $E_{\rm cut}=5$ GeV. The \mlhad decoder also reproduces well the distribution of  hadronization chain multiplicities. Only a few hadronization events result in just a few hadrons, a bulk of hadronization events contain between 7 to 13 hadrons, and both hadronization chain generators feature a tail of rather long hadronization chains. The differences between the \pythia and \mlhad hadron multiplicity distributions are in most cases at the level of $5-10\%$, where the largest deviations occur for hadronization events with just a few hadron emissions. This is expected, given that \pythia and \mlhad models of hadronization differ in the treatment of the very first emission, see the discussion at the beginning of Section~\ref{eq:sec:pythia:reproduce}.

\begin{figure}[t]
        \centering
        \includegraphics[width=0.8\textwidth]{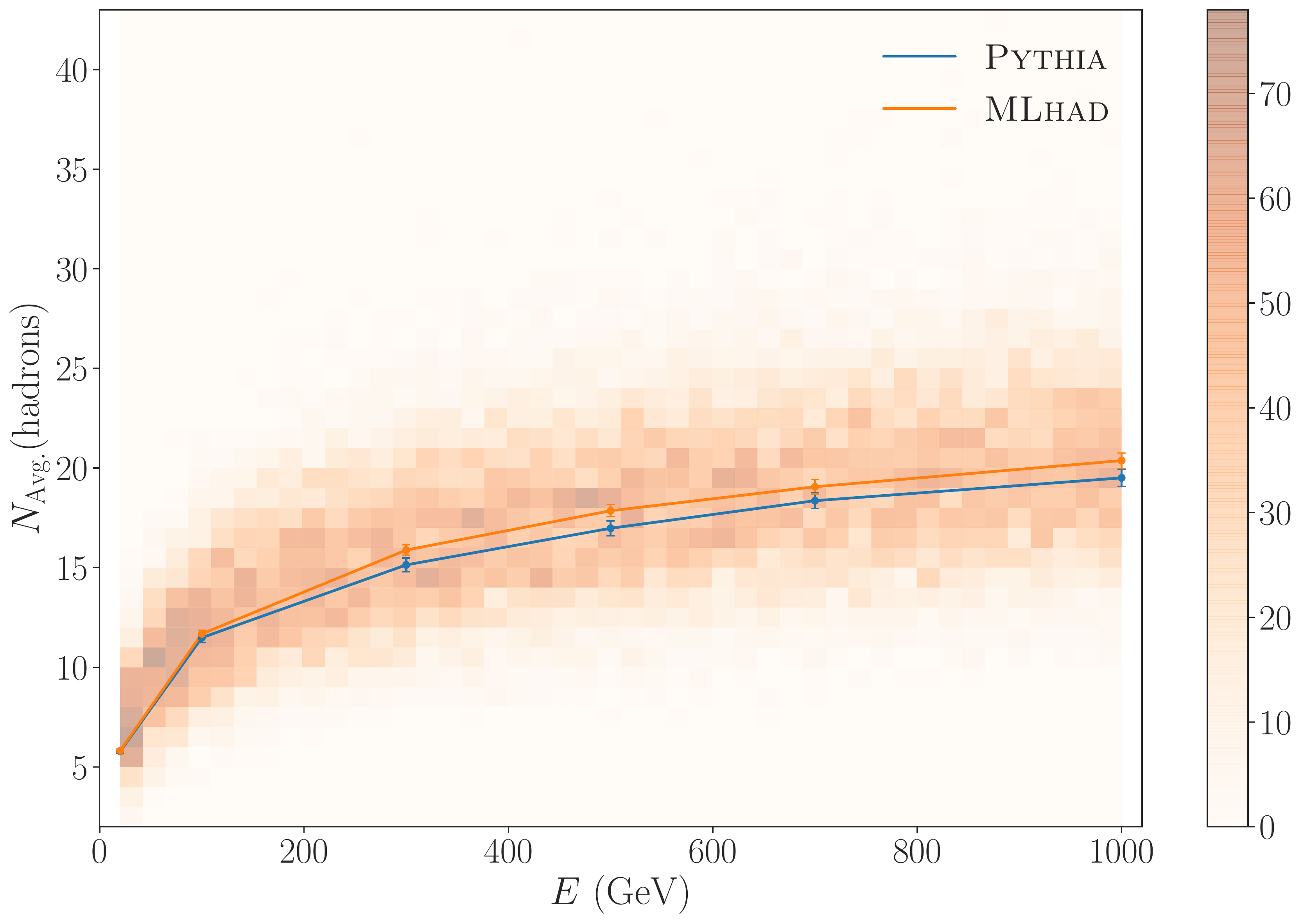}
        \caption{Comparison of the average number of hadrons produced in the fragmentation chain of a single string as a function of the initial parton energy $E$ ($E_\text{string}=2E$), produced using \pythia (blue) and \mlhad (red). The density plot shows the multiplicity distributions obtained with \mlhad for $2\times 10^4$ fragmentation chains.}
        \label{fig:figure:density}
\end{figure}

 In Fig. \ref{fig:figure:density} we also show the comparison of the average multiplicity of the hadronization chain as a function of the initial parton energy, obtained either with \pythia (blue solid line) or with \mlhad (red).  We see that \mlhad is able to reproduce the \pythia fragmentation chain length averages, and in particular also give the expected $\log E$ dependence of the average number of produced hadrons. For each energy the multiplicity distributions also match well, which we checked explicitly, while in the figure we only show the result for \mlhad to guide the eye (red density plot). The density plot scan was performed by randomly choosing an initial parton energy $E$ between $20$ GeV-$1000$ GeV and binning each fragmentation chain length with a parton energy resolution of 22 GeV and chain length resolution of 1.7 hadrons for a total of $2\times10^4$ fragmentation events. The minimal initial string energy was chosen to be 20 GeV such that it is still well above the imposed hadron emission cut $E_{\rm cut}=5$ GeV. 

\section{Conclusions}
\label{sec:conclusions}
The cSWAE architecture that was developed in this work appears to be well suited for modeling the nonperturbative process of hadronization -- the creation of hadrons from the energy stored in the string connecting a $q\bar q$ pair. We have demonstrated this by training the \mlhad hadronization models to a simplified version of \pythia hadronization, limited to only light quark flavor  endings of the string, and allowing only for pions to be the final-state hadrons. Furthermore, we utilized the scaling properties of the \pythia hadronization model that simplified the cSWAE training, requiring training at just a single string energy. Even so, the results shown in Figs. \ref{pz_gen_grid}, \ref{pT_gen_grid} and \ref{fig:figure4} are very encouraging. The \pythia first-hadron emission distributions at a fixed string energy, Fig.~\ref{pz_gen_grid}, \ref{pT_gen_grid}, are faithfully reproduced by the \mlhad decoder, as are the hadron multiplicities for full hadronization chains, Fig.~\ref{fig:figure4}.

The cSWAE architecture also has enough built in flexibility that it should be possible to extend the \mlhad model to handle all possible string flavors and kinematics. We have already shown that the inclusion of a label allows for an interpolation  of the hadronization models to different string energies, see Fig.~\ref{label_pz_gen}. This should then also allow to extend the \mlhad models below the string energy cut of 5 GeV that we imposed in this preliminary exploration. Similarly, the conditional label could  be used for  \mlhad to handle the generation of hadron flavors, including possible kinematic dependencies. The \mlhad architecture should also allow us to model any correlations between $p_z$ and $p_T$ distributions of the emitted hadrons, if these are present in data, even though currently we used the absence of such correlations in \pythia generated data to simplify the training of \mlhad models. Another important feature that we anticipate to be particularly important once \mlhad is trained directly on experimental data, is the flexibility in the choice of the latent-space distributions, making it easier to adapt to any possible features not captured by the rather constrained form of the Lund fragmentation function underlying the hadronization implementation in \pythia.
Finally, some of the planned extensions of the \mlhad hadronization framework may require more thought, most notably how to best model the hadronization of baryons and include gluons. 

 While in this paper the training of \mlhad was performed on the first hadron emissions in the \pythia output, such training will not be possible when using real experimental data, since such information is physically not  possible to extract directly from data. Instead, the training will need to be performed on the physically accessible observables constructed from particle flows measured either in $e^+e^-$ or $pp$ collisions with two, three or more  jets in the final state. We anticipate that this is where the machine learning approach to hadronization will prove most useful --- capturing the many observables in principle available in the data, such as hadron multiplicities, angular separations and momentum distributions for various hadrons. This data-collection is tedious when performed through human intervention and is a problem that calls for a machine learning based optimization. We believe that the presented \mlhad cSWAE architecture is well suited to achieve this next step.

\section*{Acknowledgments}
We thank Jared Evans for collaboration in the initial stages of this work, and Stephen Mrenna, Manuel Szewc, and Mike Williams for useful comments on the manuscript.

\paragraph{Funding information.}
AY, JZ, and TM  acknowledge support in part by the DOE grant de-sc0011784 and NSF OAC-2103889. PI is supported in part by NSF OAC-2103889.

\begin{appendix}
\section{Public code \tt{MLhad\_v0.1} }
\label{sec:App:A}
The public code may be accessed through \url{https://gitlab.com/uchep/mlhad}. The public directory includes example files allowing the user to train and implement cSWAE models in full fragmentation chains. The programs are written in \python and extensively use the \pythia, \pytorch and \scikitlearn libraries. Installation instructions can be found on the respective installation pages for each library.

The provided programs can be split into two categories: training cSWAE models and generating hadronization events. The latter relies on the former.  However, we have also provided pre-trained models such that the user can generate hadronization events without explicitly training a model.

Training a unique model configuration can be done by modifying the files \texttt{pT\_SWAE.py}, \texttt{pz\_SWAE.py}, or \texttt{pz\_cSWAE.py}. The SWAE programs contain examples of label-independent training, while the cSWAE program provides an example of label-dependent training. The model hyperparameters and target latent distribution described in Section~\ref{sec:SWAEs} have been set to default values to provide a reasonable starting configuration but may be modified. Label independent kinematic training datasets for $p_z$ and $p_T$ have been provided as well as a label-dependent $p_z$ dataset.

Full hadronization events use the trained model decoder to generate hadronic kinematics. An example of generating this kinematic data from SWAE trained model decoders can be found in \texttt{model\_pxpypz.py}. The setup of our modified fragmentation chain which utilizes these kinematics can be seen in \texttt{frag\_chain.py}.

\section{Sliced Wasserstein distance}
\label{sec:app:sliced:Wasserstein}
In this appendix we give a short overview of the Wasserstein distance and the sliced-Wasserstein distance.

\paragraph{The Wasserstein distance.} The Earth mover's distance or the Wasserstein distance gives a measure of how different two distributions are,  
given a metric space $\Omega$ and a space of Borel probability measures $\mathcal{P}(\Omega)$ on $\Omega$. The $p$-Wasserstein distance $W_p(\mu, \nu)$ between any two probability measures $\mu \in \mathcal{P}(X)$ and $\nu \in \mathcal{P}(Y) $ is \cite{villani2008}
\begin{equation}
W_p(\mu, \nu):= \bigg(\underset{\pi \in \Pi(\mu, \nu)}{\rm inf} \int_{X} c(x,y)  d\pi(x,y)   \bigg)^{\frac{1}{p}} ,
\end{equation}
where $c(x,y)$ is the cost function,  $\Pi (\mu, \nu)$ is the set of all transportation plans, with $\pi \in \Pi (\mu, \nu)$, while $p \in [1, \infty)$. The distance $W_1$ is also commonly called the Kantorovich-Rubinstein distance.

If $\mu$ and $\nu$ are one-dimensional measures, the Wasserstein distance has a closed-form expression  
\begin{equation}
\label{eq:Wp:1D}
W_p(\mu, \nu) = \bigg( \int_0^1 |F_{\mu}^{-1}(z) - F_{\nu}^{-1}(z)|^p dz \bigg)^{1/p} ,
\end{equation}
where $F_{\mu(\nu)}(x) =  \int_{- \infty }^x I_{\mu(\nu)}(\tau) d\tau$ are the cumulative distribution functions, with $I_{\mu}$ and $I_{\nu}$ the probability density functions for the measures $\mu$ and $\nu$, respectively. 
The $W_p(\mu, \nu)$ for  the one dimensional case can therefore be calculated by simply sorting the samples from the two distributions and calculating the average cost.

\paragraph{Radon transform and the sliced-Wasserstein distance.} 
An approximate value for the Wasserstein distance $W_p$ between two higher dimensional distributions on $X=\mathcal{R}^d$ can be obtained efficiently from a set of projections to one-dimensional distributions, since for each of these one can use the closed form of Eq.~\eqref{eq:Wp:1D}.
The projection from the higher dimensional distribution to the one-dimensional representation is done by the Radon transform.

The $d$-dimensional Radon transform $R$ maps a function $I \in L^1(\mathcal{R}^d)$ to \cite{Helgason2015}
\begin{equation}
R I(t, \theta) = \int_{\mathcal{R}^d} |I(x)| \delta(t - \langle x,\theta \rangle) dx ,
\end{equation}
with $(t, \theta) \in \mathcal{R} \times \mathcal{S}^{d-1}$, where $\mathcal{S}^{d-1}$ is the unit sphere in $\mathcal{R}^d$, $\delta(\cdot)$ is the delta function and $\langle , \rangle$ is the Euclidean scalar product. For a fixed direction $\theta$ the Radon transform $R I_\mu(\cdot, \theta)$ therefore gives a one dimensional marginal distribution of $I_\mu$ that is obtained by integrating $I_\mu$ over the hyperplane orthogonal to $\theta$. 

The sliced-Wasserstein distance $SW_p(I_{\mu}, I_{\nu})$ between $I_{\mu}$ and $I_{\nu}$ is defined as 
\begin{equation}
\label{eq:SWp}
SW_p(I_{\mu}, I_{\nu}) = \bigg( \int_{\mathcal{S}^{d-1}} W_p(R I_{\mu}(\cdot,\theta), R I_{\nu}(\cdot, \theta) d\theta \bigg)^{\frac{1}{p}}.
\end{equation}
The Wasserstein distance between each of the one dimensional projections (slicings)  $R I_{\mu}(\cdot,\theta)$ and $R I_{\nu}(\cdot, \theta)$ is obtained straightforwardly using the closed form result of Eq.~\eqref{eq:Wp:1D}. The integral over the unit sphere vectors $\theta$ probes all the possible slicings. Furthermore, $SW_p(I_\mu, I_\nu)$ approximates $W_p(I_\mu,I_\nu)$ ``well enough'' \cite{Santambrogio2015}.

The integration in Eq.~\eqref{eq:SWp} over the unit sphere in $\mathcal{R}^d$ can be estimated using a Monte Carlo integration that draws samples $\{\theta_l \}$ from the uniform distribution on $\mathcal{S}^{d-1}$. This replaces the integral with a finite sample average,
\begin{equation}
SW_p(I_{\mu}, I_{\nu}) \approx \bigg( \frac{1}{L} \sum_{l=1}^L W_p(R I_{\mu}(\cdot, \theta_l), R I_{\nu}(\cdot, \theta_l) )   \bigg)^{\frac{1}{p}} ,
\end{equation}
where $L$ is the number of projections (slicings). With this result, the sliced-Wasserstein distance is obtained by solving a finite number of one-dimensional optimal transport problems, each of which has a closed-form solution. Furthermore, the sliced-Wasserstein distance approximates well the Wasserstein distance and thus can be used as a useful discriminator for the similarity of distributions.  More details can be found in \cite{DBLP:journals/corr/abs-1902-00434} and \cite{DBLP:journals/corr/abs-1804-01947}.

\section{Latent distributions}\label{latent_dist}
The analytic forms of the latent target distributions used in the training of cSWAE in Section~\ref{eq:trained:models} are
\begin{equation}
    I_{\text{tri.}}(z;a,b,c) = 
    \begin{dcases}
      \frac{2(z-a)}{(b-a)(c-a)}, &  a \leq z \leq c ,\\
      \frac{2(b-z)}{(b-a)(b-c)}, &  c < z \leq b, \\
    \end{dcases}
\eeq
for the triangular distribution, and
\beq
    I_{\text{trap.}}(z;a,b,c,d) = 
    \begin{dcases}
      \frac{2}{d+c-a-b}\frac{z-a}{b-a}, & a \leq z < b, \\
      \frac{2}{d+c-a-b}, & b \leq z < c, \\
      \frac{2}{d+c-a-b}\frac{d-z}{d-c}, &  c \leq z \leq d,
    \end{dcases}
\end{equation}  
for the trapezoidal distribution. For a given initial parton energy $E$ the choices of parameters $a,b,c,d$ can be seen in Table~\ref{tab:pz_fit_params}. The target latent-space distributions are then given by 
\beq
\label{eq:latent:I}
I_{\text{tri.}}(\bz,\bc) =\prod_{k=1}^{N_e} I_{\text{tri.}}(z_k;a,b,c), \qquad I_{\text{trap.}}(\bz,\bc) =\prod_{k=1}^{N_e} I_{\text{trap}}(z_k;a,b,c,d), 
\eeq
that is we take the same values of $a,b,c,d$ parameters for all $d_z$ latent dimensions. 

The normal and skewed-normal distributions are given by 
\begin{align}
  I_{\text{Gauss}}(z;\mu, \sigma) = \frac{1}{\sigma \sqrt{2\pi}} \exp \left( -\frac{(z - \mu)^2}{2\sigma^2} \right),
\\
  I_{\text{Skew-Gauss}}(z; \mu, \sigma, \alpha) = 2 I_{\text{Gauss}}(z; \mu, \sigma) \Phi\left( \frac{\alpha(z - \mu)}{\sigma} \right),
\end{align}
respectively, where
\begin{equation}
  \Phi(x) = \frac{1}{\sqrt{2\pi}} \int_{-\infty}^x e^{-t^2/2}dt.
\end{equation}
The $\mu$, $\sigma$, and $\alpha$ are the fit parameters corresponding to the mean, standard deviation, and skewness of the distribution, respectively. As in Eq.~\eqref{eq:latent:I} the $d_z$ dimensional latent-space distributions are products of one dimensional ones with the same $\mu, \sigma, \alpha$ parameters. For $p_T$ we have $\mu=0.099$, $\sigma=0.257$, and $\alpha=4.259$.

\begin{table}[t]
\centering
\begin{tabular}{ c | c | c  c  c  c }
\toprule
  Variable $\boldsymbol{x}$ & Target $\boldsymbol{z}$ & $a$ & $b$ & $c$ & $d$ \\
 \midrule
 \multirow{2}{*}{$p_z'$} & Trapezoidal & $0.04 E$ & $0.16 E$ & $0.24E$ & $E$ \\
  & Triangular & $0.04 E$ & $0.2 E$ & $E$ & -- \\
$p_T$ & Triangular & 0.0 & 0.3 & 1.0 & -- \\
\bottomrule
\end{tabular}
\caption{The $p_z'$ and $p_T$ latent-space distribution parameters.}
\label{tab:pz_fit_params}
\end{table}

\end{appendix}

\bibliography{VAE_hadronization.bib}

\begin{thebibliography}{10}
\providecommand{\url}[1]{\texttt{#1}}
\providecommand{\urlprefix}{URL }
\expandafter\ifx\csname urlstyle\endcsname\relax
  \providecommand{\doi}[1]{doi:\discretionary{}{}{}#1}\else
  \providecommand{\doi}{doi:\discretionary{}{}{}\begingroup
  \urlstyle{rm}\Url}\fi
\providecommand{\eprint}[2][]{\url{#2}}

\bibitem{Alwall:2014hca}
J.~Alwall, R.~Frederix, S.~Frixione, V.~Hirschi, F.~Maltoni, O.~Mattelaer,
  H.~S. Shao, T.~Stelzer, P.~Torrielli and M.~Zaro,
\newblock \emph{{The automated computation of tree-level and next-to-leading
  order differential cross sections, and their matching to parton shower
  simulations}},
\newblock JHEP \textbf{07}, 079 (2014),
\newblock \doi{10.1007/JHEP07(2014)079},
\newblock \eprint{1405.0301}.

\bibitem{Sjostrand:2014zea}
T.~Sj\"ostrand, S.~Ask, J.~R. Christiansen, R.~Corke, N.~Desai, P.~Ilten,
  S.~Mrenna, S.~Prestel, C.~O. Rasmussen and P.~Z. Skands,
\newblock \emph{{An introduction to PYTHIA 8.2}},
\newblock Comput. Phys. Commun. \textbf{191}, 159 (2015),
\newblock \doi{10.1016/j.cpc.2015.01.024},
\newblock \eprint{1410.3012}.

\bibitem{Bellm:2015jjp}
J.~Bellm \emph{et~al.},
\newblock \emph{{Herwig 7.0/Herwig++ 3.0 release note}},
\newblock Eur. Phys. J. C \textbf{76}(4), 196 (2016),
\newblock \doi{10.1140/epjc/s10052-016-4018-8},
\newblock \eprint{1512.01178}.

\bibitem{Bothmann:2019yzt}
E.~Bothmann \emph{et~al.},
\newblock \emph{{Event Generation with Sherpa 2.2}},
\newblock SciPost Phys. \textbf{7}(3), 034 (2019),
\newblock \doi{10.21468/SciPostPhys.7.3.034},
\newblock \eprint{1905.09127}.

\bibitem{Andersson:1983ia}
B.~Andersson, G.~Gustafson, G.~Ingelman and T.~Sjostrand,
\newblock \emph{{Parton Fragmentation and String Dynamics}},
\newblock Phys. Rept. \textbf{97}, 31 (1983),
\newblock \doi{10.1016/0370-1573(83)90080-7}.

\bibitem{Andersson:1998tv}
B.~Andersson,
\newblock \emph{{The Lund model}},
\newblock Camb. Monogr. Part. Phys. Nucl. Phys. Cosmol. \textbf{7}, 1 (1997).

\bibitem{Ferreres-Sole:2018vgo}
S.~Ferreres-Sol\'e and T.~Sj\"ostrand,
\newblock \emph{{The space\textendash{}time structure of hadronization in the
  Lund model}},
\newblock Eur. Phys. J. C \textbf{78}(11), 983 (2018),
\newblock \doi{10.1140/epjc/s10052-018-6459-8},
\newblock \eprint{1808.04619}.

\bibitem{Field:1982dg}
R.~D. Field and S.~Wolfram,
\newblock \emph{{A QCD Model for e+ e- Annihilation}},
\newblock Nucl. Phys. B \textbf{213}, 65 (1983),
\newblock \doi{10.1016/0550-3213(83)90175-X}.

\bibitem{Gottschalk:1983fm}
T.~D. Gottschalk,
\newblock \emph{{An Improved Description of Hadronization in the \{QCD\}
  Cluster Model for $e^+ e^-$ Annihilation}},
\newblock Nucl. Phys. B \textbf{239}, 349 (1984),
\newblock \doi{10.1016/0550-3213(84)90253-0}.

\bibitem{Webber:1983if}
B.~Webber,
\newblock \emph{{A QCD Model for Jet Fragmentation Including Soft Gluon
  Interference}},
\newblock Nucl. Phys. B \textbf{238}, 492 (1984),
\newblock \doi{10.1016/0550-3213(84)90333-X}.

\bibitem{Bishara:2019iwh}
F.~Bishara and M.~Montull,
\newblock \emph{{(Machine) Learning amplitudes for faster event generation}}
  (2019),
\newblock \eprint{1912.11055}.

\bibitem{Badger:2020uow}
S.~Badger and J.~Bullock,
\newblock \emph{{Using neural networks for efficient evaluation of high
  multiplicity scattering amplitudes}},
\newblock JHEP \textbf{06}, 114 (2020),
\newblock \doi{10.1007/JHEP06(2020)114},
\newblock \eprint{2002.07516}.

\bibitem{Gao:2020vdv}
C.~Gao, J.~Isaacson and C.~Krause,
\newblock \emph{{i-flow: High-dimensional Integration and Sampling with
  Normalizing Flows}},
\newblock Mach. Learn. Sci. Tech. \textbf{1}(4), 045023 (2020),
\newblock \doi{10.1088/2632-2153/abab62},
\newblock \eprint{2001.05486}.

\bibitem{Gao:2020zvv}
C.~Gao, S.~H\"oche, J.~Isaacson, C.~Krause and H.~Schulz,
\newblock \emph{{Event Generation with Normalizing Flows}},
\newblock Phys. Rev. D \textbf{101}(7), 076002 (2020),
\newblock \doi{10.1103/PhysRevD.101.076002},
\newblock \eprint{2001.10028}.

\bibitem{Chahrour:2021eiv}
I.~Chahrour and J.~D. Wells,
\newblock \emph{{Function Approximation for High-Energy Physics: Comparing
  Machine Learning and Interpolation Methods}}  (2021),
\newblock \eprint{2111.14788}.

\bibitem{Winterhalder:2021ngy}
R.~Winterhalder, V.~Magerya, E.~Villa, S.~P. Jones, M.~Kerner, A.~Butter,
  G.~Heinrich and T.~Plehn,
\newblock \emph{{Targeting Multi-Loop Integrals with Neural Networks}}  (2021),
\newblock \eprint{2112.09145}.

\bibitem{Matchev:2020tbw}
K.~T. Matchev, A.~Roman and P.~Shyamsundar,
\newblock \emph{{Uncertainties associated with GAN-generated datasets in high
  energy physics}}  (2020),
\newblock \eprint{2002.06307}.

\bibitem{Alanazi:2020klf}
Y.~Alanazi \emph{et~al.},
\newblock \emph{{Simulation of electron-proton scattering events by a
  Feature-Augmented and Transformed Generative Adversarial Network (FAT-GAN)}}
  (2020),
\newblock \doi{10.24963/ijcai.2021/293},
\newblock \eprint{2001.11103}.

\bibitem{Nachman:2020fff}
B.~Nachman and J.~Thaler,
\newblock \emph{{Neural resampler for Monte Carlo reweighting with preserved
  uncertainties}},
\newblock Phys. Rev. D \textbf{102}(7), 076004 (2020),
\newblock \doi{10.1103/PhysRevD.102.076004},
\newblock \eprint{2007.11586}.

\bibitem{Stienen:2020gns}
B.~Stienen and R.~Verheyen,
\newblock \emph{{Phase space sampling and inference from weighted events with
  autoregressive flows}},
\newblock SciPost Phys. \textbf{10}(2), 038 (2021),
\newblock \doi{10.21468/SciPostPhys.10.2.038},
\newblock \eprint{2011.13445}.

\bibitem{Butter:2020qhk}
A.~Butter, S.~Diefenbacher, G.~Kasieczka, B.~Nachman and T.~Plehn,
\newblock \emph{{GANplifying event samples}},
\newblock SciPost Phys. \textbf{10}(6), 139 (2021),
\newblock \doi{10.21468/SciPostPhys.10.6.139},
\newblock \eprint{2008.06545}.

\bibitem{Backes:2020vka}
M.~Backes, A.~Butter, T.~Plehn and R.~Winterhalder,
\newblock \emph{{How to GAN Event Unweighting}},
\newblock SciPost Phys. \textbf{10}(4), 089 (2021),
\newblock \doi{10.21468/SciPostPhys.10.4.089},
\newblock \eprint{2012.07873}.

\bibitem{Danziger:2021eeg}
K.~Danziger, T.~Jan\ss{}en, S.~Schumann and F.~Siegert,
\newblock \emph{{Accelerating Monte Carlo event generation -- rejection
  sampling using neural network event-weight estimates}}  (2021),
\newblock \eprint{2109.11964}.

\bibitem{Butter:2021csz}
A.~Butter, T.~Heimel, S.~Hummerich, T.~Krebs, T.~Plehn, A.~Rousselot and
  S.~Vent,
\newblock \emph{{Generative Networks for Precision Enthusiasts}}  (2021),
\newblock \eprint{2110.13632}.

\bibitem{Biro:2021zgm}
G.~B\'\i{}r\'o, B.~Tank\'o-Bartalis and G.~G. Barnaf\"oldi,
\newblock \emph{{Studying Hadronization by Machine Learning Techniques}}
  (2021),
\newblock \eprint{2111.15655}.

\bibitem{Howard:2021pos}
J.~N. Howard, S.~Mandt, D.~Whiteson and Y.~Yang,
\newblock \emph{{Foundations of a Fast, Data-Driven, Machine-Learned
  Simulator}}  (2021),
\newblock \eprint{2101.08944}.

\bibitem{Quetant:2021hgi}
G.~Qu\'etant, M.~Drozdova, V.~Kinakh, T.~Golling and S.~Voloshynovskiy,
\newblock \emph{{Turbo-Sim: a generalised generative model with a physical
  latent space}}  (2021),
\newblock \eprint{2112.10629}.

\bibitem{Bieringer:2022cbs}
S.~Bieringer, A.~Butter, S.~Diefenbacher, E.~Eren, F.~Gaede, D.~Hundhausen,
  G.~Kasieczka, B.~Nachman, T.~Plehn and M.~Trabs,
\newblock \emph{{Calomplification -- The Power of Generative Calorimeter
  Models}}  (2022),
\newblock \eprint{2202.07352}.

\bibitem{Buhmann:2020pmy}
E.~Buhmann, S.~Diefenbacher, E.~Eren, F.~Gaede, G.~Kasieczka, A.~Korol and
  K.~Kr\"uger,
\newblock \emph{{Getting High: High Fidelity Simulation of High Granularity
  Calorimeters with High Speed}},
\newblock Comput. Softw. Big Sci. \textbf{5}(1), 13 (2021),
\newblock \doi{10.1007/s41781-021-00056-0},
\newblock \eprint{2005.05334}.

\bibitem{Ilten:2016csi}
P.~Ilten, M.~Williams and Y.~Yang,
\newblock \emph{{Event generator tuning using Bayesian optimization}},
\newblock JINST \textbf{12}(04), P04028 (2017),
\newblock \doi{10.1088/1748-0221/12/04/P04028},
\newblock \eprint{1610.08328}.

\bibitem{Andreassen:2019nnm}
A.~Andreassen and B.~Nachman,
\newblock \emph{{Neural Networks for Full Phase-space Reweighting and Parameter
  Tuning}},
\newblock Phys. Rev. D \textbf{101}(9), 091901 (2020),
\newblock \doi{10.1103/PhysRevD.101.091901},
\newblock \eprint{1907.08209}.

\bibitem{radford2016unsupervised}
A.~Radford, L.~Metz and S.~Chintala,
\newblock \emph{Unsupervised representation learning with deep convolutional
  generative adversarial networks} (2016), \eprint{1511.06434}.

\bibitem{kingma2014autoencoding}
D.~P. Kingma and M.~Welling,
\newblock \emph{Auto-encoding variational bayes} (2014), \eprint{1312.6114}.

\bibitem{DBLP:journals/corr/abs-1804-01947}
S.~Kolouri, C.~E. Martin and G.~K. Rohde,
\newblock \emph{Sliced-wasserstein autoencoder: An embarrassingly simple
  generative model},
\newblock CoRR \textbf{abs/1804.01947} (2018),
\newblock \eprint{1804.01947}.

\bibitem{DBLP:journals/corr/abs-1711-01558}
I.~O. Tolstikhin, O.~Bousquet, S.~Gelly and B.~Sch{\"{o}}lkopf,
\newblock \emph{Wasserstein auto-encoders},
\newblock CoRR \textbf{abs/1711.01558} (2017),
\newblock \eprint{1711.01558}.

\bibitem{NEURIPS2019_9015}
A.~Paszke, S.~Gross, F.~Massa, A.~Lerer, J.~Bradbury, G.~Chanan, T.~Killeen,
  Z.~Lin, N.~Gimelshein, L.~Antiga, A.~Desmaison, A.~Kopf \emph{et~al.},
\newblock \emph{Pytorch: An imperative style, high-performance deep learning
  library},
\newblock In H.~Wallach, H.~Larochelle, A.~Beygelzimer, F.~d'Alch\'{e} Buc,
  E.~Fox and R.~Garnett, eds., \emph{Advances in Neural Information Processing
  Systems 32}, pp. 8024--8035. Curran Associates, Inc. (2019).

\bibitem{villani2008}
C.~e. Villani,
\newblock \emph{Optimal transport, old and new},
\newblock Springer, Berlin (2008).

\bibitem{Helgason2015}
S.~Helgason,
\newblock \emph{Integral Geometry and Radon Transforms},
\newblock Springer, New York (2015).

\bibitem{Santambrogio2015}
F.~Santambrogio,
\newblock \emph{Optimal Transport for Applied Mathematicians},
\newblock Springer, Switzerland (2015).

\bibitem{DBLP:journals/corr/abs-1902-00434}
S.~Kolouri, K.~Nadjahi, U.~Simsekli, R.~Badeau and G.~K. Rohde,
\newblock \emph{Generalized sliced wasserstein distances},
\newblock CoRR \textbf{abs/1902.00434} (2019),
\newblock \eprint{1902.00434}.

\end{thebibliography}

\nolinenumbers

\end{document}